\documentclass[a4paper,fleqn]{cas-dc}

\usepackage[authoryear]{natbib}

\begin{document}

\let\WriteBookmarks\relax
\def\floatpagepagefraction{1}
\def\textpagefraction{.001}
\shorttitle{Ionosphere and solar evolution}
\shortauthors{K. Mursula}

\title [mode = title]{Centennial solar EUV irradiance from ionospheric currents: Varying sunspot-EUV irradiance relation and modified spot-facula ratio}                      

%
%

\author[1]{K. Mursula}[
                        ]

\ead{kalevi.mursula@oulu.fi}
\ead[url]{https://www.oulu.fi/en/research-groups/space-climate}


\address[1]{Space Climate Group, Space Physics and Astronomy Res. Unit, University of Oulu, Finland}

\begin{abstract}
Sunspots offer a uniquely long view of solar magnetic activity, and depict large variability during the last 100 years, a period known as the Modern Maximum (MM).
However, if our view of solar variability was only based on the strongest magnetic fields, it would be incomplete.
Therefore, other variables are needed to study the long-term evolution, e.g., of weaker fields and different radiative emissions.
Recently, the relation between sunspots and several other solar activity proxies like the F10.7 and F30 radio fluxes and the MgII index (proxies of EUV irradiance) was found to vary during the last 70 years so that a relative sunspot dominance over EUV in the 1950s-1960s changed to EUV dominance in the 2000s \citep{Mursula_AA_2024}.
Here we use data from eight long-operating observatories to calculate the yearly range of daily variation of the geomagnetic Y-component, the rY index, for the last 137 years.
We show that the rY index correlates extremely well with the MgII index and the solar F30 radio flux. 
These three indices also have no trend relative to each other over the respective intervals.
On the other hand, the F10.7 flux has a significant trend with respect to the three co-varying EUV indices (MgII, F30, rY).
Therefore, the rY index replaces F10.7 as the best long-term EUV proxy, and extends the MgII index by 90 years.
We verify that all the four EUV proxies (rY, MgII, F30, F10.7) have an increasing trend with respect to sunspots during the last 50-70 years.
This is valid both for sunspot numbers and group numbers. 
Extending this earlier with the rY index, we find that the relation between EUV irradiance and sunspots has a nonlinear, quadratic  evolution over the MM.
This implies that the Sun has more sunspots relative to EUV irradiance during the growth and maximum of the MM, while the opposite is true during its decay.
We estimate that the MgII index (solar EUV irradiance) increases by 24\% of its solar cycle variation with respect to the sunspot number during the last 70 years.
Our results indicate a systematic difference in the evolution between sunspots (generally: photosphere) and plages (generally: chromosphere) with long-term solar activity.
The implied varying spot-facula ratio has consequences to the stellar evolution of the Sun and Sun-like stars.

\end{abstract}




 \maketitle

%
\section{Introduction}
	\label{sec:Introduction}

Sunspots are the visible imprint of the strongest magnetic fields on the solar surface produced by the dynamo mechanism operating in the solar convection layer \citep[for recent reviews, see, e.g., ][]{Cameron_2017, Charbonneau_2020, Stenflo2015}.
Sunspots vax and wane on the solar surface, the photosphere, according to the roughly 11-year solar cycle.
Moreover, sunspot cycle heights have greatly varied during the 400 years of telescopic monitoring, depicting a near-centennial Gleissberg cyclicity \citep{Gleissberg_1939, Ogurtsov_2002, Feynman_2014, Hathaway_LRSP_2015} during the last three centuries after the Maunder minimum
\citep{Spoerer_1887, Maunder_Spoerer_MNRAS_1890, Eddy_1976, Usoskin_2015}.

The latest Gleissberg cycle occurred during the 20th century when the Sun experienced a period of unprecedented activity, now called the Modern Maximum (MM) \citep{Usoskin_2003, Solanki_2004, Hathaway_LRSP_2015}. 
Sunspot cycle (SC) heights varied from the low cycles at the turn of the 19th and 20th centuries to the MM maximum during SC19 in 1950s and back roughly to the same low level during SC24 in 2010s as one century ago.
Sunspot number (SSN version 2) reduced from its all-time maximum of 285.0 in SC19 to 116.4 at the maximum of SC24.
Accordingly, sunspot cycle heights reduced by a factor of about 2.4 during the decay of the MM.

However, if our view of solar variability was only based on sunspots, the strongest magnetic fields which represent only a small and very specific form of solar magnetism, it would remain one-sided and, in fact, incomplete.
This is emphasised by the fact that the surface magnetic field is by far dominated by weak magnetic fields, much weaker than the sunspots \citep[see, e.g., ][]{deWijn_2009, Rutten_2021}.
Therefore, other variables, independent of sunspot numbers, are needed to study the long-term evolution of weaker fields and, e.g., of the different solar radiative emissions.

Because the relative brevity of space age compared to the era of sunspot observations, satellite observations of the Sun are typically limited to the last 50 years, 
These include, e.g., the solar EUV irradiance measurements, like to MgII core-to-wing ratio index that has been measured since 1978.
Luckily there are a few long, homogeneous series of ground-based observations of the Sun that can be used to study other aspects of solar magnetism.
Perhaps the longest and most reliable parameters are the solar radio flux measurements that have been conducted in Canada at 10.7\,cm (F10.7) wavelength since the late 1940s and in Japan at different wavelengths (e.g., 30\,cm; F30) since the late 1950s.
Note that the solar radio fluxes are considered and used as close proxies of solar EUV irradiance, especially at monthly and longer time scales \citep{Tapping_2013, Schonfeld_2019}.

A number of studies have examined the mutual relation between sunspots and solar 10.7\,cm radio flux, finding that this relation changed from 1970s until 2010s so that the Sun emitted more of radio flux relative to sunspots in the last two decades of this period than in the earlier decades \citep{Tapping_2011, Tapping_2017, Bruevich_2019, Lastovicka_SpW_2023, Mursula_AA_2024}.
Accordingly, sunspot activity decreased relatively faster than EUV irradiance when solar activity was weakening during the decay of the Modern Maximum.
Recently, we have verified that a similar change occurred in the relation between sunspots and several EUV proxies (F10.7, F30 and MgII) during the last 70 years, so that a relative sunspot dominance over EUV in the 1950s-1960s changed to relative EUV dominance over sunspots in the 2000s \citep{Mursula_AA_2024}.

The above results are limited to the last 70 years when solar activity was, in the beginning, very high (MM maximum) and then declining during the decay of the Modern Maximum.
This, as such, is a rather specific situation, which may lead to questionable or, at least, partial results.
Therefore, it is highly desirable to extend these studies in time and to learn how these relations developed, e.g., around the turn of centuries more than 100 years ago and thereafter, in the growth phase of the MM.
We have recently presented preliminary results for these early times since the 1880s \citep{Mursula_AA_2025} using the early measurements of the Earth's magnetic field and a very old method based on the dependence of dayside ionospheric currents on solar EUV irradiance.

\cite{Graham_1724} studied the Earth's magnetic field and found out almost exactly 300 years ago that the declination of the geomagnetic field experiences a systematic daily variation with a maximum in the morning and minimum in the afternoon. 
Graham had no explanation to his observation but it is now known that this variation reflects the magnetic effect of the dayside electric current system called the $S_q$ current produced by the combined effect of the ionisation of the Earth's upper atmosphere by solar EUV irradiance and the Earth's rotation \citep[for a review, see ][]{Yamazaki_2017}).

In the early 1850s J. von Lamont \citep{Lamont_1851}, 
J.-A. Gautier \citep{Gautier_1852} and R. Wolf \citep{Wolf_1852} noted that the amplitude of the daily variation of declination follows the sunspot cycle, which was found somewhat earlier by S.H. Schwabe \citep{Schwabe_1844}.
When constructing his renowned sunspot numbers, Wolf noted that there is a highly significant linear relation between the yearly sunspot number and the yearly averaged declination amplitude \citep{Wolf_1859}.
He also used this relation to predict the declination amplitude from sunspots and, vice versa, to fill in data gaps in his sunspot series using the declination amplitude.

It took several decades before the chain of physical processes producing the correlation between sunspots and declination amplitude was disclosed. 
A central idea to explain the correlation between sunspots and declination amplitude, first suggested by \cite{Stewart_1882}, was that the upper atmosphere is electrically conducting and a site of large, persistent electric currents which can cause magnetic effects on the ground and produce a diurnally varying declination.
\cite{Schuster_1908} proposed that these currents are driven by solar extreme ultraviolet (EUV) radiation which ionises the dayside upper atmosphere.
However, satellite observations were needed to prove that solar EUV radiation indeed varies in phase with sunspot cycle \citep{Hinteregger_1979, Donnelly_1986}.
This finally gave the motivation for Wolf's linear correlation between sunspots and declination amplitude, more than a century after the initial finding.

Here we extend our preliminary analysis by calculating the daily range of the geomagnetic East (Y)-component (rY index) at eight early and long-running stations with homogeneous measurements to develop a centennial (137-year long) proxy of solar EUV irradiance and to study its relation with sunspots over the whole Modern Maximum.
The paper is organised as follows. 
We introduce the geomagnetic data and and solar indices in Section \ref{sec:Data}, and study the daily variation of the magnetic Y-component in Section \ref{sec:Daily variation}.
Section \ref{sec:Secular variation} examines the centennial (secular) variation at the eight stations and compares it to daily variation.
In Section \ref{sec:Yearly rY} we calculate the yearly rY indices, and in Section \ref{sec:8st_mean_rY} we combine them to form multi-station rY indices, which we compare with sunspot number in Section \ref{sec:3means_rY_SSN}.
In section \ref{sec:Long_rY} we construct two different 130-year long rY indices and compare them with sunspot number in Section \ref{sec:Long_SSN}.
In Section \ref{sec:Long_rY_GSN} we compare the official long rY series with three versions of group sunspot numbers, and in Section \ref{sec:Long_rY_EUV} with three (other) indices of EUV irradiance (MgII, F10.7, and F30).
In Section \ref{sec:Discussion} we discuss our results and present their implications to evolution of the Sun during the Modern Maximum and to the stellar evolution of the Sun and Sun-like stars.
Finally, Section \ref{Sec: Conclusions} presents our conclusions.

%
\section{Data}
\label{sec:Data}

\subsection{Geomagnetic stations}

We use here hourly measurements of the Y (East-West) -component of the geomagnetic field made at eight long-operating or early magnetic observatories: Cheltenham - Fredericksburg series (IAGA station code: CLH/FRD; here called CLH), Ekaterinburg (EKT), Cape Town - Hermanus series (CTO/HER; here called HER), Honolulu (HON), Kakioka (KAK), Niemegk (NGK), San Juan (SJG) and Tucson (TUC).
Table \ref{table:Stations} shows the IAGA codes, the geographic coordinates (latitude and longitude), the UT hours of local midnight, and the start and end years of operations of these stations.
All the used magnetic data can be obtained from the portal of the World Data Center for Geomagnetism (Edinburgh) maintained by the British Geological Survey (https://wdc-dataportal.bgs.ac.uk).
We have imposed the condition that no longer than 2-month long data gaps are allowed in the Y-component data in any year. 
Otherwise the data of that year are omitted (set to NaN). 
The omitted years are also denoted in Table \ref{table:Stations} for each station.

The selected observatories are among the most reliable stations, each of them producing verified, homogeneous data for several decades, some longer than one century.
Some of them are standard stations producing internationally well known and largely used geomagnetic indices. 
For example, data from HER, HON, KAK, SJG are used to calculate the storm-time Dst index \citep{Sugiura_1964} 
and its extension, the Dxt index \citep{Karinen_Mursula_2005}.
NGK is the base station for all K-indices, including the Kp/Ap indices of global geomagnetic activity \citep{Matzka_2021}.
Continuous observations started at the NGK station (actually at its predecessor Potsdam) already in 1890.
EKT is part of a Russian net of magnetic stations, the earliest of which belonged to the famous magnetic crusade initiated by C. F. von Gauss. 
EKT has the earliest digitised data among the stations selected to this study, starting in 1887.
Alas, EKT data only extend until 1925. 
Luckily this is sufficiently long so that meaningful comparisons can be made with the other observatories.
We note that the EKT Y component in the WDC database had a sign error and the LT time was erroneously given as the UT time. 
We have noted the WDC on these errors. 
However, these flaws, even if uncorrected, would have no effect on the daily range that we use here.


\begin{table}[width=.99\linewidth,cols=6,pos=h]
\caption{IAGA codes, geographic latitudes and longitudes, midnight UT hours, and start, end and omitted years for the eight magnetic stations.}             
\label{table:Stations}   
\begin{tabular*}{\tblwidth}{@{} LLLLLLL@{} }
\toprule 		
Station & GGlat & GGLong & MiUT & Start & End & Omitted\\
\midrule  		 
\textbf{CLH}&38.73&1-76.84&5&1902&2023&1901\\
\textbf{EKT}&56.83&60.63&20&1887&1925&1923 \\
\textbf{HER}&-34.43&19.23&23&1933&2023&1932\\
\textbf{HON}&21.32&-158&11&1902&2023&1978\\
\textbf{KAK}&36.23&140.18&15&1913&2023&1917-1923\\
\textbf{NGK}&52.07&12.68&23&1890&2023&1945-1947\\
\textbf{SJG}&18.38&-66.12&4&1926&2023&1927,2017,2018\\
\textbf{TUC}&32.17&-110.73&7&1910&2023&1909\\
\bottomrule
\end{tabular*}
\end{table}

\subsection{Solar parameters}

We use here several long-term measures of solar activity, the Wolf/International sunspot number (SSN) series, three different series of group sunspot numbers (GSN), 
two solar radio fluxes at 10.7\,cm (F10.7) and 30\,cm (F30) wavelength and the Bremen MgII index.
Sunspot numbers and group sunspot numbers give a view of the evolution of photospheric magnetic activity, while the radio fluxes are proxies of solar EUV irradiance, and measure solar activity at two slightly different altitudes of the chromosphere and lower corona, and the MgII index is a standard measure of EUV irradiance.
 
For SSN we use the yearly sunspot numbers of version 2 \citep{Clette_2015, Clette_Preface_2016}.
For GSN we use, firstly, the collection of sunspot groups from 1610 to 2010 by \cite{Vaquero_2016} (to be called here VAQ GSN), which is a revision of the original collection of groups by \cite{Hoyt_SP2_1998}, secondly, the collection of sunspot groups from 1610 to 2015 by \cite{Svalgaard_2016} (SVA GSN) and, thirdly, the collection of sunspot groups from 1739 to 2010 by Chatzistergos \cite{Chatzistergos_2017} (CHA GSN).
Both the SSN series and all the three GSN series are served at the World Data Center SILSO of the Royal Observatory of Belgium (https://www.sidc.be/silso/).
We note further that our results will remain essentially the same for SSN version 1, as well as for some other GSN collections that have been proposed.

Continuous measurements of the solar 10.7\,cm radio flux have been made in Canada since 1947 \citep{Covington_1947, Tapping_2013}.
The solar radio F10.7 index consists of  hourly means of radio-wave flux density over a frequency band of about 100\,MHz centered at 2800\,MHz, given in terms of solar radio flux units (1\,sfu = $10^{-22}$Wm$^{-2}$Hz$^{-1}$). 
The Canadian measurements were used by the NOAA to construct the most used version of the daily F10.7 index from 1947 until the end of April, 2018, when the NOAA stopped the index production.
We extracted the daily NOAA F10.7 indices from the Lisird database (https://lasp.colorado.edu/ lisird/) and use them until April 2018.
We then continued the NOAA F10.7 series from May 2018 onward using the more recent 10.7\,cm flux measurements available from the Canadian NRCan server, as described in more detail by \cite{Mursula_AA_2023}.
This produces a homogeneous F10.7 index from 1947 onward, which is used in this study.

Solar radio flux observations in Japan started in the 1950s at four different wavelengths (for a recent review, see \cite{Shimojo_2023}). 
Measurements were first, since 1951, made in Toyakawa with 3.75 GHz (8 cm) waves.
Later in the 1950s they also started measurements at three other frequencies (1, 2, and 9.4\,GHz; corresponding wavelengths 30\,cm, 15\,cm, and 3.2\,cm). 
Observations were made in Toyakawa until 1994 and continued thereafter in Nobeyama by the National Astronomical Observatory of Japan, together with three additional bands of higher frequency.
We use here the 30\,cm flux (1\,GHz) measurements, which are available from March 1957 onward, and can be retrieved from the Lisird database. 
(Note that only data with flag = 0 are used here to guarantee good data quality).

  \begin{figure*}  
   \centering
    \includegraphics[width=0.98\linewidth]{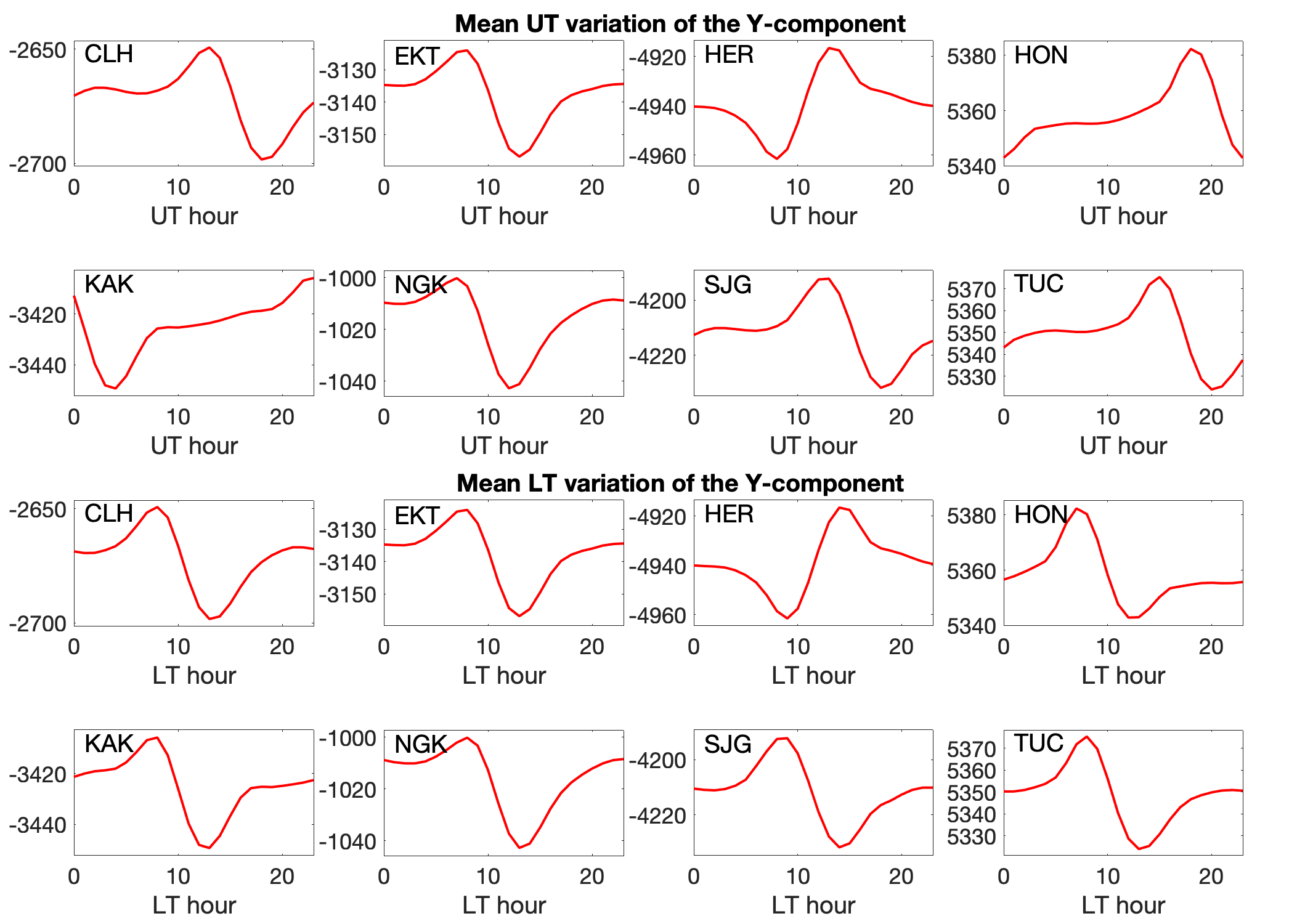}
    \caption{The daily variation of the Y-component at the eight stations, each in its own panel. 
    Top two rows: Variation in the UT time. 
    Bottom two rows: Variation in the local time.} 
   \label{fig:8st_UT_LT_daily_curves_EKT_corr}
\end{figure*}

The core to wing ratio of the Magnesium-II doublet at 280\,nm (MgII index) is a standard measure for solar UV-EUV irradiance \citep[see, e.g.,][]{DeLand_1993, Viereck_1999, Viereck_2001, Viereck_2004, Snow_2014}.
Magnesium-II measurements from several satellites (GOME, SCIAMACHY, GOME-2A, and GOME-2B) have been used to construct a long-term Mg II index, the so-called Bremen MgII composite index
updated by M. Weber (https://www.iup. uni-bremen.de/gome/gomemgii.html). 
The MgII index is an index of the overall chromospheric activity and, as will be discussed later in more detail, has a close connection to solar plages.
We will use the Bremen MgII composite index as the definite measure of solar EUV activity and compare its long-term variation with that of several other solar and geomagnetic-based variables.

%
\section{Daily variation of geomagnetic Y-component}
\label{sec:Daily variation}

The two top (bottom) rows of Fig. \ref{fig:8st_UT_LT_daily_curves_EKT_corr} depict the mean daily variation of the Y-component in the UT time (in the local time, LT, respectively) at the eight selected stations over their full respective operation periods.
When the Y-component is depicted in UT time (two top rows of Fig. \ref{fig:8st_UT_LT_daily_curves_EKT_corr}), the variations are quite different, except for those stations, like CLH and SJG which are roughly in the same longitude and hemisphere. 
However, when local time is used (two bottom rows of Fig. \ref{fig:8st_UT_LT_daily_curves_EKT_corr}), all the seven stations in the northern hemisphere depict a closely similar variation with a morning maximum and afternoon minimum, while the southern station (HER) shows an opposite variation with a morning minimum and afternoon maximum.

The daily LT time variation of the Y-component reflects the magnetic effect of the dayside electric current system called the $S_q$ (solar quiet) or the $S_R$ (solar regular) current, which is formed in the equatorial ionospheric E-layer by the combined effect of the ionisation of the Earth's upper atmosphere by solar EUV irradiance and the Earth's rotation (for a review, see \citet{Yamazaki_2017}).
The $S_q$ current system consists of two current vortices, one on either hemispheric side of the sub-solar point.
While the $S_q$-current is oriented mainly along the Y-axis around the noon, the current diverges in the north-south direction in the morning and afternoon, thus causing there a deflection in the magnetic Y-component. 
Since the Y-component determines the local declination (deviation from the north), the $S_q$-current leads to a regular daily variation of declination, which was found by the early observers of the geomagnetic field \cite{Graham_1724}.
Because the $S_q$ current makes oppositely running vortices in the two hemispheres, the deflections in the morning and afternoon are opposite in the northern and southern hemisphere.
Note also that, although the daily curves in the LT time are very similar, their absolute levels are very different and can be either positive or negative, depending on the orientation of the magnetic field at the location of the station.

%
\section{Secular evolution}
\label{sec:Secular variation}

  \begin{figure*}		
   \centering
  \includegraphics[width=0.98\linewidth]{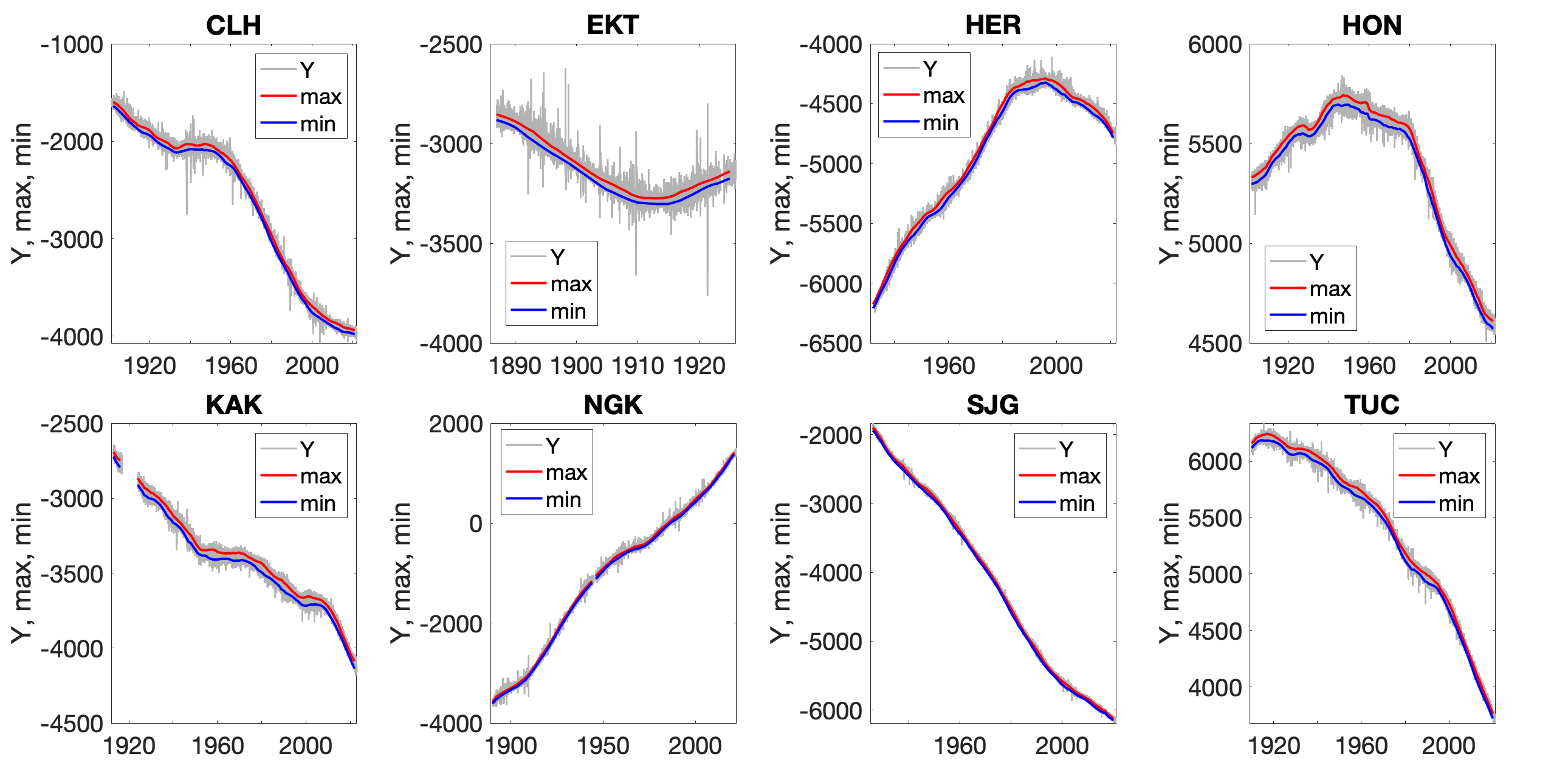}
    \caption{
 Secular variation of the Y-component at the eight stations at hourly resolution (grey line), and yearly means of the daily maxima (red curve) and minima (blue curve) of the Y-component of each station. Unit of the y-axis is nanotesla (nT). 
 } 
 \label{fig:secular_curves_Y_max_min_8st}
\end{figure*}

The eight panels of Fig. \ref{fig:secular_curves_Y_max_min_8st} depict the multi-decadal to centennial (secular) variation of the Y-component of the eight stations at hourly resolution (grey line), as well as the yearly means of the daily maxima (red curve) and minima (blue curve) of the Y-component of each station. 
The form, scale and the range of variation of the Y-component are very different at the eight stations, depending on the location of the station and how the geomagnetic field evolves at the corresponding station site. 
The range of secular variation varies from the largest range of more than 4000\,nT at SJG to the smallest secular range of about 1200\,nT at HON. 
(At EKT it is only some 350\,nT but the time series is too short to determine the actual secular variation).

The coloured lines of the yearly means of daily maxima and minima run close to each other over the whole time interval in each panel, indicating that their difference, i.e., the typical yearly range of daily variation is quite constant and much smaller than the range of secular variation.
The separation (and visibility) between the daily maximum and minimum curves is somewhat larger for those stations, like EKT and HON, where the range of secular variation is small than for those stations, like NGK and SJG, where it is large. 
This indicates that the typical range of daily variation is quite similar at all stations, despite the very different ranges of secular  variation.
Indeed, as we will see later in Sec. \ref{sec:Yearly rY}, the yearly ranges of daily variation are, on an average some 40-50\,nT for each station. 

On the other hand, the ratio of the range of daily variation to the range of secular variation varies from about 10\% to less than 1\% in the different stations. 
This shows that the secular variation has little or no effect on the long-term evolution of the range of daily variation.
Note also that the hourly values of the Y-component (grey line in Fig. \ref{fig:secular_curves_Y_max_min_8st}) depict some rather large peak-like variations.
These are due to magnetic storms causing variations of up to several hundred nT, i.e., much larger than the typical range of daily variation, mainly in the night local time sector.

%
\section{Yearly rY indices }
\label{sec:Yearly rY}

\begin{figure*}  
   \centering
  \includegraphics[width=0.98\linewidth]{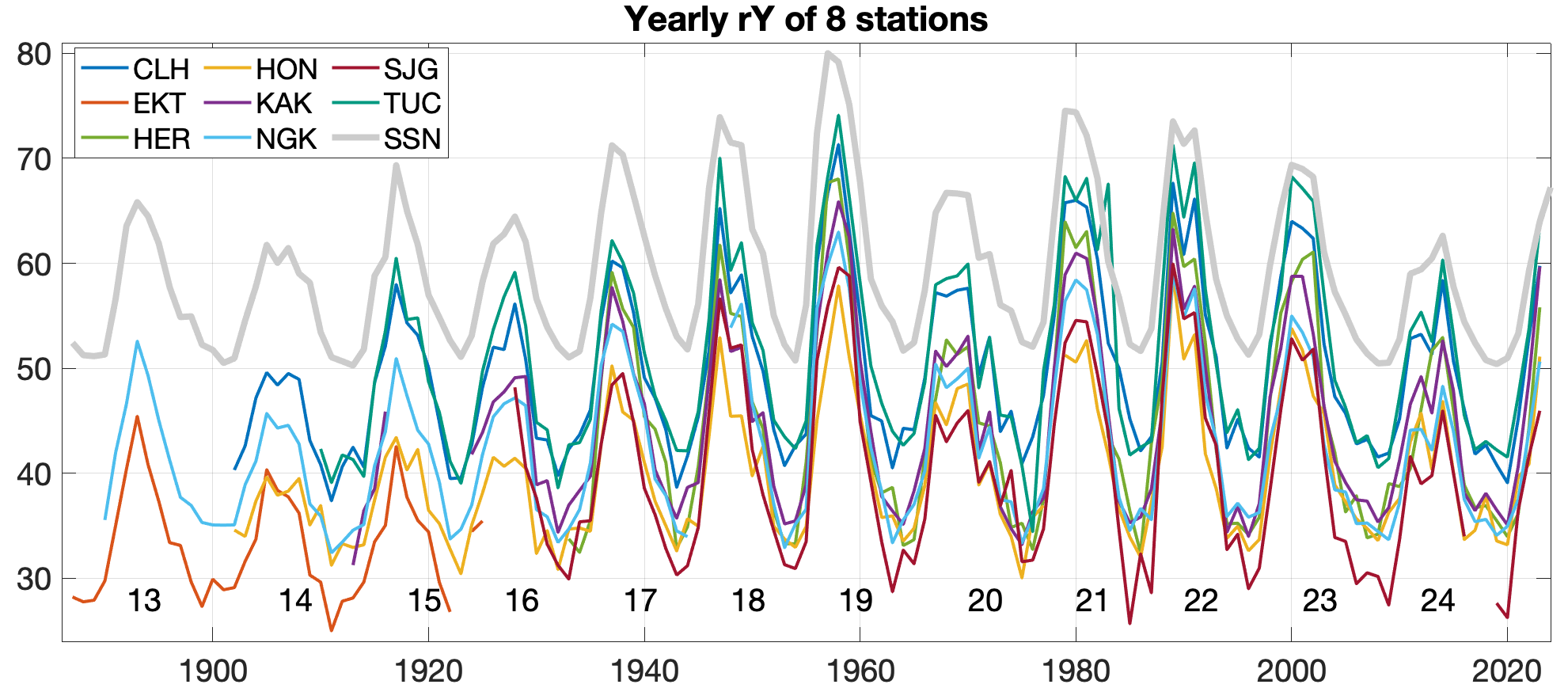}
    \caption{Yearly rY indices for the 8 stations over their full time range (colored lines).
    Scaled sunspot numbers are included for comparison (grey line).} 
   \label{fig:rY_yearly_8stations_full}
\end{figure*}

The difference between the daily maximum and minimum of the Y-component, i.e., the daily range of the Y-component is also called the (daily) rY index \citep{Svalgaard_EUV_2016, Mursula_AA_2025}. 
We have calculated yearly averages of the daily rY indices for the eight stations and depicted them in Fig. \ref{fig:rY_yearly_8stations_full}.
(Here we use all available days for the calculation of a yearly average since the daily variation is practically the same for quiet and all days, as first noted by \cite{Chree_1913}).

Figure \ref{fig:rY_yearly_8stations_full} verifies the above noted fact that the yearly rY ranges of the eight stations are quite similar (on an average about 40-50\,nT) although the absolute level and the range of secular variation of the Y-component are very different at the eight stations. 
EKT and SJG are at a slightly lower overall level and TUC at a slightly higher level than other stations, but the differences are indeed very small compared to the differences in their  secular variations.

Figure \ref{fig:rY_yearly_8stations_full} shows that the rY indices of all stations depict a cyclic evolution, closely following the sunspot cycles (included in Fig. \ref{fig:rY_yearly_8stations_full} for comparison).
The rY index cycles of all stations even reproduce the centennial variation of sunspot cycle heights over the Modern Maximum (MM), the latest Gleissberg cycle \citep{Gleissberg_1939}, with the highest rY cycle found, for each station (except for EKT, of course), during the highest sunspot cycle 19, which forms the MM maximum.

%
\section{Multi-station mean rY indices}
\label{sec:8st_mean_rY}

We will now form three multi-station mean rY indices by combining the rY indices of a number of stations that have been operating during a common time interval.
Combining one-station rY indices together to multi-station mean indices will give an even more systematic evolution of the rY values over the common time with less random variation than in any one station alone.
Moreover, plotting the station-based rY indices together with the multi-station mean will allow us to see the possible differences and the level of similarity between the yearly rY values from the different stations during the common time of operation.

Table \ref{table:Stations} shows that all other stations except for EKT have been in operation in most years from 1933 to 2023.
This is the temporal extent of one of the three multi-station mean rY indices to be called the 7-station (7-st) rY index.
The 5-station (5-st) mean rY consists of the rY indices of the 5 longest-running stations (CLH, HON, KAK, NGK, TUC), extending the 5-station mean to start in 1913, i.e., 20 years earlier than the 7-station rY index, covering 111 years.
Finally, the two earliest stations (EKT, NGK) are combined to a 2-station (2-st) mean rY index over their common time from 1890 to 1925.

When combining the stations to a multi-station mean rY, each one-station rY series has to be standardised over the respective common time interval in order to set all the stations to the same level and in order to guarantee that all stations have an equal weight to the mean.
Standardisation removes the mean from the rY series and divides it by its standard deviation.
Therefore, the standardised values fluctuate around zero and the unit of the standardised y-axis is the standard deviation.
Accordingly, when calculating each of the three multi-station mean rY series, we have standardised the rY series of all the contributing stations and then averaged them to form the multi-station mean rY.
(Note that for the 7-st and 5-st means, each of the contributing stations has to be standardised separately for the two means because the time intervals are different.)

\begin{figure*}  
   \centering
  \includegraphics[width=0.98\linewidth]{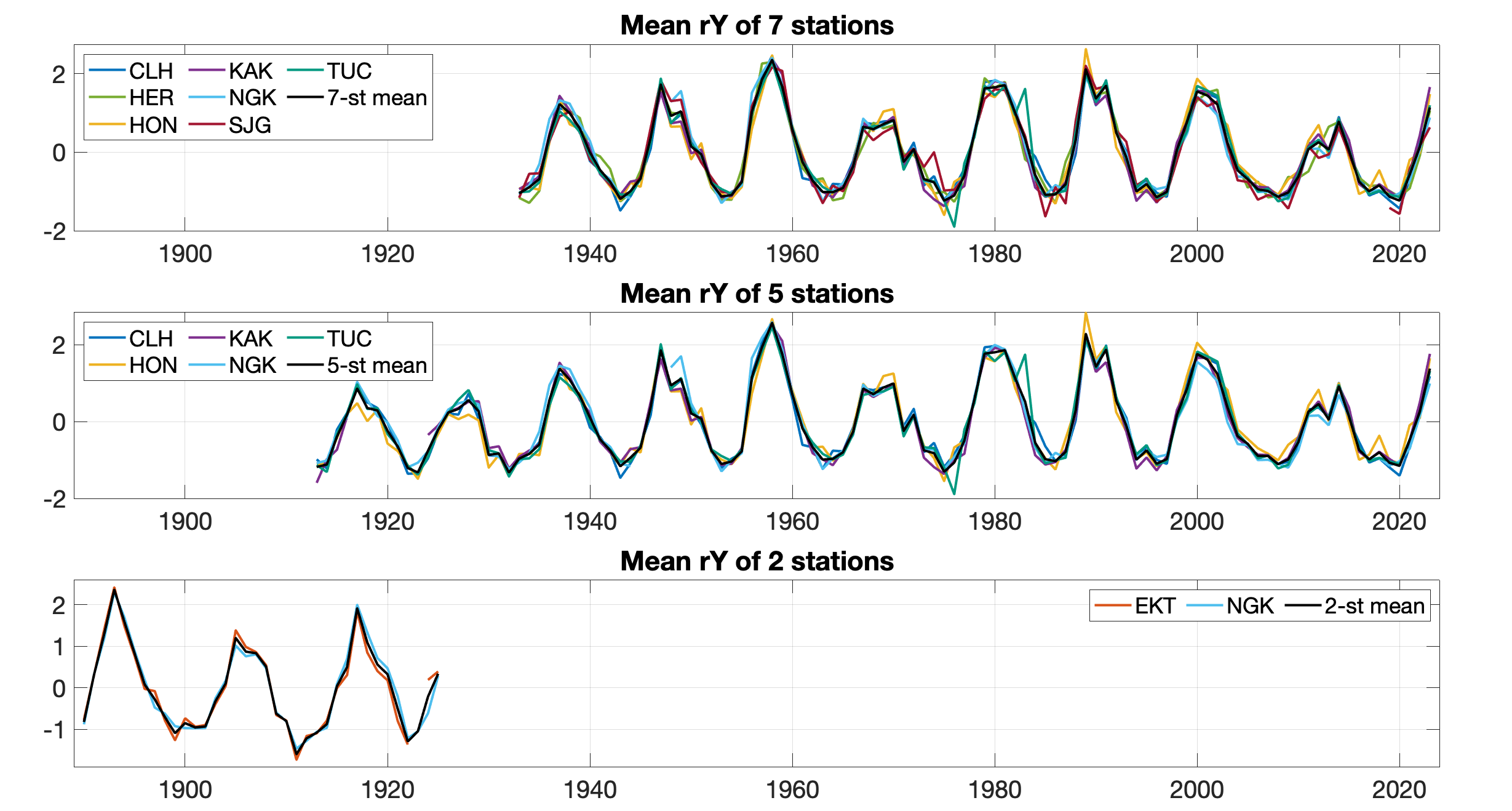}
    \caption{Yearly rY values of the three multi-station means  (black line) and the contributing station rY values (colored lines) constructed from the 8 stations.
    Top: The 7-station mean and the contributing stations in 1933-2023. 
    Middle: The 5-station mean and the contributing stations in 1913-2023.
    Bottom: The 2-station mean and the contributing stations in 1890-1925.
   } 
   \label{fig:rY_3means_8stations}
\end{figure*}

Figure \ref{fig:rY_3means_8stations} shows the three  (7-st, 5-st and 2-st) multi-station mean rY indices together with the standardised rY indices of the contributing stations.
One can see that the rY indices of all the contributing stations follow very closely to the respective multi-station mean rY index in each of the three cases. 
This is particularly true for the 2-station mean (bottom panel of Fig. \ref{fig:rY_3means_8stations}) where one can see hardly any difference between the two contributing stations (EKT and NGK).
The mean absolute deviation of the two stations from the 2-station mean is only 0.10\,$\sigma$ (units of st.dev.).
Compared to the average amplitude of the 2-st mean rY (and the two one-station rY's) of about 3\,$\sigma$, this gives a relative accuracy of cycle amplitude determination of about 0.1/3 $\approx$ 3\%.

In the 7-station mean, one can see occasional deviations of some stations from the mean for one or two successive years.
By far the largest yearly absolute deviation of about 1.28\,$\sigma$ is in TUC rY in 1983, as clearly seen in Fig.  \ref{fig:rY_3means_8stations}.
The second largest absolute deviation of about 0.79\,$\sigma$ is also in TUC rY (in 1976). 
Despite these two years, TUC rY values are, overall, closest to the 7-st mean of all the 7 stations, deviating from it (in absolute value) only by 0.108\,$\sigma$ on an average.
This shows how the multi-station mean can correct the individual erroneous values of the contributing stations.

Figure \ref{fig:rY_3means_8stations} shows also some other, smaller deviations from the mean.
The third largest absolute deviation of about 0.77\,$\sigma$ is in SJG rY in 1974.
All other yearly deviations of all 7 stations are clearly smaller, with the average absolute deviation being 0.136\,$\sigma$.
Overall, HER (0.172), HON (0.174) and SJG (0.171) have somewhat larger average absolute deviations than CLH (0.112), KAK (0.116), NGK (0.125) and TUC (0.108). 
Note that HON and SJG had omitted years in the middle of data, while KAK and NGK had one long gap and for CLH, HER and TUC only the start year was omitted (see Table \ref{table:Stations}).
Concluding, the data gaps affect slightly to data quality but their overall effect to the 7-station mean remains very small.

Two of the three most-deviating stations (HER and SJG) are absent in the 5-station mean, which reduces the overall average absolute deviation to 0.126\,$\sigma$, some 0.01\,$\sigma$ lower than in the 7-st mean.
Therefore, the 5-station and 7-station means have an incredible correlation (in the overlapping time of 1933-2023) with cc = 0.999 and p =  $1.7\cdot 10^{-116}$.
Moreover, the mean absolute difference between the two rY mean series is only 0.084, i.e., less than the mean absolute deviation of any one-station rY index from either multi-station mean.
This shows that the positive and negative deviations of one-station rY values largely cancel out to the overall mean, leading to very similar multi-station means irrespective of the number (7 or 5) of contributing stations.

Note also that, although all of the one-station rY indices deviate from the multi-station mean, occasionally even by a considerable amount, none of them systematically deviate from the mean.
Most importantly, since standardisation does not affect the trend of the index, this means that the rY indices of all the seven, five or two stations, as well as their respective 7-st, 5-st and 2-st means, evolve mutually closely similarly and have closely similar long-term trends over the whole respective time interval (1933-2023, 1913-2023, 1890-1925).

%
\section{Correlating multi-station rY with SSN}
\label{sec:3means_rY_SSN}

\begin{figure*}  
   \centering
  \includegraphics[width=0.95\linewidth]{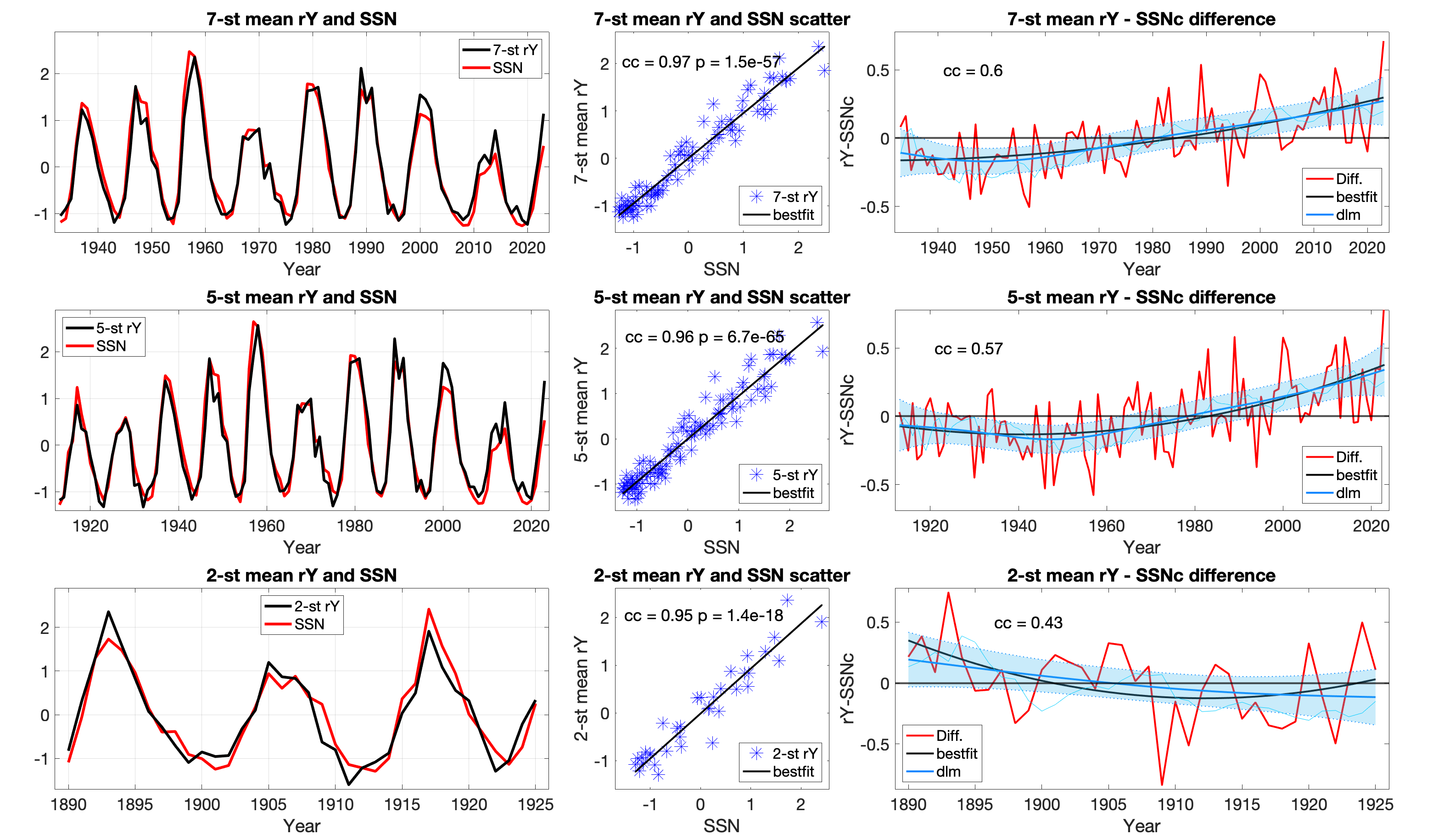}
    \caption{Correlating the three multi-station mean rY indices with sunspot number.
    Top row: 7-station mean rY; Middle row: 5-station mean rY; Bottom row: 2-station mean rY.
    Left column: Multi-station mean rY and sunspot number time series;
    Middle column: Their scatterplot with best-fit line;
    Right column: Difference (residual) between the multi-station mean rY indices and the correlated sunspot number, together with the best-fit 2nd-order polynomial and the dynamic linear model.
   } 
   \label{fig:rY_3means_8stations_SSN_relation}
\end{figure*}

The main aim of this study is to examine the long-term relation between sunspot activity and EUV activity using various rY indices.
For that aim, we have standardised the sunspot number (SSN) over the time intervals covered by the three multi-station mean rY indices and correlated the standardised SSN with each of the three  multi-station rY indices.
The top row of Fig. \ref{fig:rY_3means_8stations_SSN_relation} shows the results from correlating the 7-station mean with SSN, the middle row the corresponding results for the 5-station mean and the bottom row for the 2-station mean.

The first column of Fig. \ref{fig:rY_3means_8stations_SSN_relation} shows that the rY cycles and sunspot cycles have roughly equal standardised amplitudes.
This means that a change of one unit of standard deviation in sunspot activity (about 71, 68 and 46 for the three cases) produces, in each case, a change of roughly one unit of standard deviation in the mean rY.
The scatterplots depicted in the middle column of Fig. \ref{fig:rY_3means_8stations_SSN_relation} show an excellent linear correlation between the three rY indices and the corresponding sunspot numbers.
The correlation coefficients, p-values and the slopes and intercepts of the best-fit lines are depicted in Table \ref{table:3mean_SSN}.


\begin{table}[width=.99\linewidth,cols=6,pos=h]
\caption{Correlation coefficients, p-values, and the best-fit line slopes and intercepts for the correlation between the three mean rY indices and SSN.}             
\label{table:3mean_SSN}   	
\begin{tabular*}{\tblwidth}{@{} LLLLLLL@{} }
\toprule 		
Mean rY & cc & p-value & slope & intercept \\
\midrule  		 
\textbf{7-st}&0.972&$1.5\cdot 10^{-57}$&0.954&$1.32\cdot 10^{-4}$\\
\textbf{5-st}&0.964&$6.7\cdot 10^{-65}$&0.948&$3.88\cdot 10^{-4}$\\
\textbf{2-st}&0.949&$1.4\cdot 10^{e-18}$&0.941&$-1.45\cdot 10^{-2}$\\
\bottomrule
\end{tabular*}
\end{table}

Correlation coefficient of the 7-st mean (0.972) is slightly larger than in the other two cases but the p-value is clearly smallest for the 5-st mean ($6.7\cdot 10^{-65}$) because of its long temporal extent. 
The good correlation in each case implies that more than 90\% of the variation of yearly mean rY indices is explained by sunspot variability. 
This and the very small p-values leave no doubt of the statistical significance of the correlation between mean rY indices and SSN.
Note also that the intercepts of the best-fit lines are very small, practically zero, which is in agreement with the estimated explaining capability and further verifies the direct driving of rY variability by solar activity.

Despite this good correlation, a detailed analysis of the time series of rY indices and the corresponding sunspot numbers (left column of Fig. \ref{fig:rY_3means_8stations_SSN_relation}) reveals small but systematic differences in the long-term trend between the rY indices and sunspot numbers.
Most maxima of the first several cycles (SC15-SC21) of both the 7-st and the 5-st mean rY indices are slightly lower than the peaks of the corresponding sunspot cycles. 
Even more clearly, the opposite is the case for the last four solar cycles (SC22-SC25), where both 7-st and 5-st rY cycle peaks are all higher than sunspot peaks.
Moreover, the rY indices around the two last sunspot minima are clearly higher than the corresponding sunspot activity.

The fact that sunspot numbers are indeed relatively higher than the 5-st rY indices in the beginning of the time series (SC15) is verified independently by the mutual relation between the 2-st rY index and the corresponding sunspot number (low left panel of Fig.  \ref{fig:rY_3means_8stations_SSN_relation}).
The 2-st rY index overlaps with the 5-st index during cycle 15 where both 2-st and 5-st rY indices are lower than the corresponding sunspot numbers.
However, interestingly, an opposite relation with relatively higher rY cycles is observed during the first two cycles (SC13-SC14) of the time interval of the 2-st rY index.

The relation between the three multi-station mean rY indices and the corresponding sunspot numbers is studied further in the right column of Fig. \ref{fig:rY_3means_8stations_SSN_relation}, which depicts the rY-SSNc difference (residual) between the rY index and the corresponding correlated sunspot number (SSNc) obtained from the correlation shown in the middle column of Fig. \ref{fig:rY_3means_8stations_SSN_relation}.

The residuals for both the 7-st index (top right panel) and the 5-st rY index (middle right panel) are mostly negative from the start until the late 1970s and mostly positive thereafter.
The slope of a linear fit to the 7-st residual ($5.11 \pm 0.76 \cdot 10^{-3}$; not shown) is definitely significantly nonzero (positive) with p = $ 1.4 \cdot 10^{-9}$. 
Correlation coefficient is cc = 0.58, explaining some 34\% of variability.
(Note that the explaining power of the mere trend remains moderate because of the rather large short-term variation of the yearly differences.)
The same applies for the 5-st rY index residual (slope  = $4.09 \pm 0.67 \cdot 10^{-3}$, cc = 0.51, p  =  $ 1.3 \cdot 10^{-8}$) although the slope is smaller and statistical significance, despite the longer time series, slightly weaker because of the nonlinear evolution of the early part of the 5-st residual.

Accordingly, the rY index has an overall increasing trend with respect to sunspot number, especially during the last 6-7 solar cycles from the 1950s onward.
This quantifies the observations made above when discussing the rY cycle and sunspot cycle heights depicted in the left column of Fig. \ref{fig:rY_3means_8stations_SSN_relation}.
This also supports the earlier results of the long-term relation between sunspot activity and several solar EUV proxies \citep{Mursula_AA_2024, Mursula_AA_2025}, as well as between sunspots EUV-driven ionospheric variables like foF2 and foE indices \citep{Lastovicka_2019,Lastovicka_SpW_2023}.

However, the 2-st rY residual (low-right panel of Fig. \ref{fig:rY_3means_8stations_SSN_relation}) indicates an overall declining trend with mostly positive values from the start until the mid-1900s, and mostly negative values thereafter.
A linear fit to the 2-st residual yields a negative slope ($-9.11 \pm 4.85 \cdot 10^{-3}$) which, however, cannot be considered statistically significant because of the rather large p = $0.069$. 
Also the correlation coefficient of the linear fit is rather small cc = 0.31, explaining only some 9.4\% of variability.

  \begin{figure*}  
   \centering
  \includegraphics[width=0.98\linewidth]{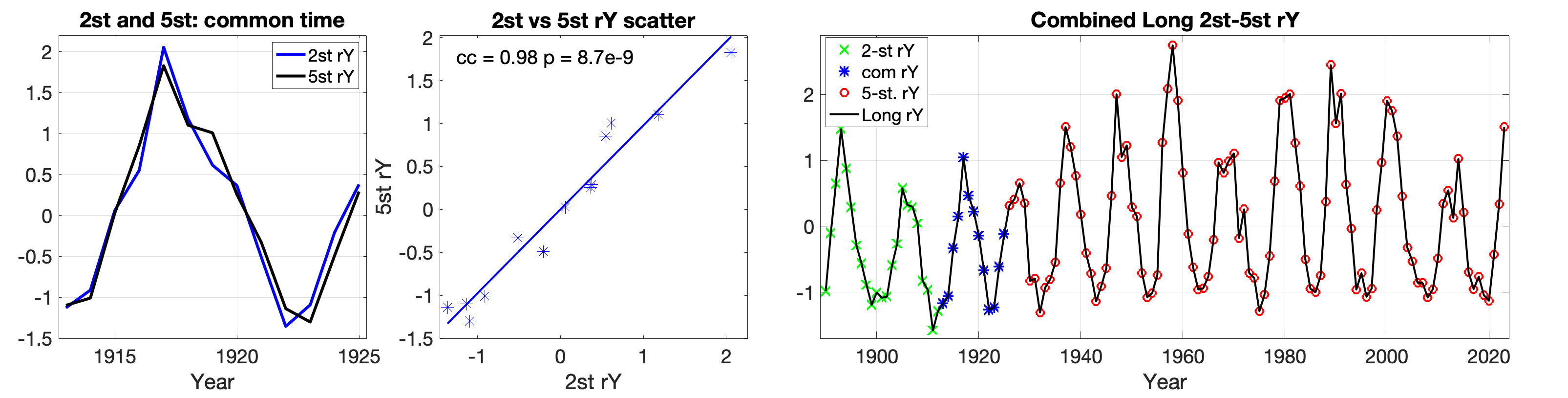}
    \caption{The 2-station and 5-station mean rY indices combined to a long rY series. 
    Left panel: 2-st and 5-st mean rY indices standardised over the common time; 
    Middle panel: Their correlation and the best-fit line; 
    Right panel: Combined long ('official') rY index (black line) consisting of three constituent parts, the 2-st rY index before the common time (green crosses), the average of 2-st and 5-st rY indices (blue stars) over the common time and the 5-st rY index after the common time (red circles).
 } 
   \label{fig:Long_2st_5st_rY}
\end{figure*}

The panels of the right column of Fig. \ref{fig:rY_3means_8stations_SSN_relation} also include the best-fit 2nd-order polynomials and the local trend estimate (blue line) and its 95\% error (blue shaded area) using the dynamic linear model (dlm; \cite{Laine_2014}), which allows the linear regression coefficient to vary in time.
A 2nd-order fit to the 2-st residual gives a larger correlation coefficient (cc =  0.43) than the linear fit, indicating for a nonlinear evolution. 
The explaining power of the 2nd-order fit, about 18.5\%, is considerably larger than that of the linear fit.
However, the 2nd-order coefficient ($-9.58 \pm 5.04 \cdot 10^{-4}$) of the best-fit polynomial to the 2-st residual cannot be considered significantly different from zero (p = 0.066).
It is likely that the weak significance of these fits to the 2-st residual is due to its much shorter time interval than in the two other cases.

For the 7-st residual the 2nd-order polynomial (cc = 0.60) and the linear fit (cc = 0.58) give roughly equally good fits.
Although the 2nd-order coefficient of the best-fit polynomial to the 7-st residual is not significantly different from zero (p = 0.12), both the 2nd-order polynomial and the dlm model suggest that the fit is rather flat in the first 2-3 decades with the increase starting only in the 1950s.

For the 5-st residual the 2nd-order polynomial gives a somewhat better fit (cc = 0.57) than the linear fit (cc = 0.51). 
Most importantly, for the 5-st residual the 2nd-order coefficient ($7.63 \pm 2.21 \cdot 10^{-5}$) of the best-fit polynomial is significantly different from zero (p = $8.2\cdot 10^{-4}$).
This verifies the nonlinear temporal evolution of the 5-st residual, and the nonlinear relation between the rY index and sunspots in the early to mid-20th century, better reached by the longer extent of the 5-st rY.

%
\section{Long rY series}
\label{sec:Long_rY}

In this Section we will construct two long-term rY series that extend over the whole 130-year time interval covered by the eight stations.
There are, obviously, a few different options to combine the rY series of the eight stations to form long rY series either using the above derived multi-station means or some other combinations of one-station rY series.
One of the two long-term series to be constructed here is a combination of two multi-station rY series, and we regard this as our best ('official') estimate for the long rY series.
Since the NGK station alone covers the whole 130-year time interval, we decided to construct the second long-term series from the other stations, not using NGK at all. 
With these choices we will have three long rY indices (official, non-NGK and NGK) that are based on quite a different selection of stations.

\subsection{Official multi-station long rY index}

While the 7-st mean rY series does not overlap with the 2-st mean, the 5-st and the 2-st mean have 13 years of common time, which covers most of sunspot cycle 15 and a part of cycle 16.
Accordingly, we will use this overlapping interval to join the two multi-station means to a long rY series to be called here the 2st-5st (or 'official') long rY index.

The left panel of Fig. \ref{fig:Long_2st_5st_rY} shows the 2-station and 5-station mean rY indices which have been standardised over the common time 1913-1925.
The two rY means follow each other very closely, with some differences in individual years so that, e.g., the cycle minima are located in successive years (1922 and 1923 for 2-st and 5-st, respectively).
Their excellent linear correlation (cc = 0.98; p = $8.7\cdot 10^{-9}$) is depicted more quantitatively in the middle panel of Fig. \ref{fig:Long_2st_5st_rY}.

  \begin{figure*}  
   \centering
  \includegraphics[width=0.98\linewidth]{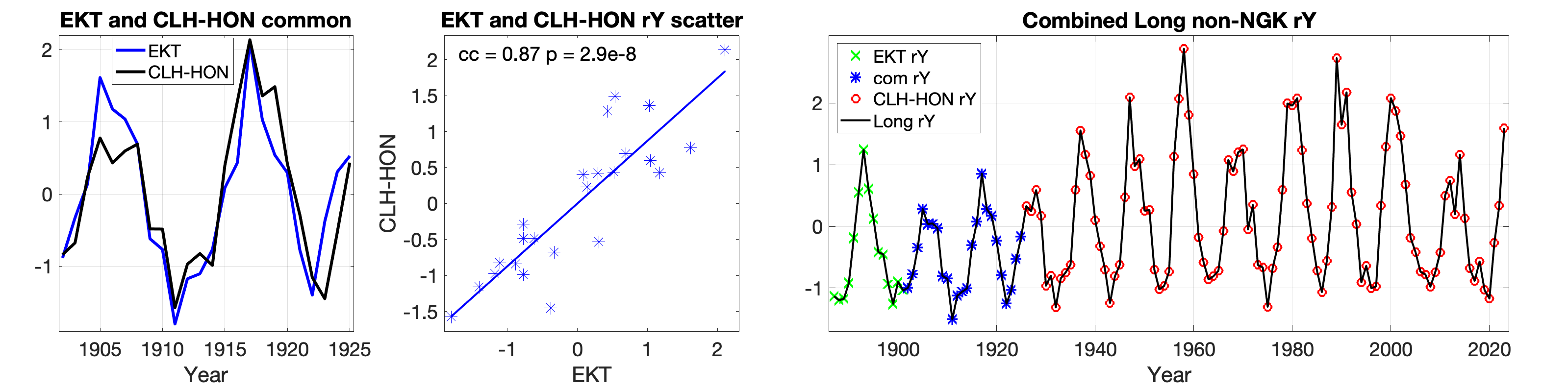}
    \caption{EKT rY index and CLH-HON mean rY index combined to a non-NGK long rY index. 
    Left panel: EKT rY and CLH-HON mean standardised over the common time; 
    Middle panel: Their correlation and the best-fit line; 
    Right panel: Combined non-NGK long rY series (black line) consisting of three constituent parts, EKT rY before the common time (blue crosses), the average of standardised EKT and CLH-HON mean rY indices (green pluses) and the CLH-HON rY index after the common time (red stars).
    } 
   \label{fig:Long_nonNGK_rY}
\end{figure*}

The right panel of Fig. \ref{fig:Long_2st_5st_rY} shows the 2st-5st ('official') long rY index consisting of three constituent parts, the 2-station rY index before the common time (green crosses), the average of the 2-station and 5-station rY indices during their common time (blue stars) and the 5-station rY index after the common time (red circles).
Before joining the three constituent series together to a long series, we have set the first part (2-station rY index before common time) and third part (5-station rY index after common time) to the same level with the respective indices in the common time.
Note that, since the multi-station mean indices were already standardised over their respective times, these level changes are not large but still necessary in order to obtain a long rY series that is homogeneous over its full extent.
Moreover, after joining, the long rY series was standardised over its full extent (1890-2023) in view of the future correlations with solar variables.

 \subsection{Non-NGK long rY index}

Here we will use stations other than NGK to construct a long rY index to be called the non-NGK long rY index.
Obviously, the first decades are then only covered by the EKT rY index. 
In addition to EKT, we will use rY indices from two longest-running stations after NGK, CLH and HON which both extend from 1902 to 2023 (see Table \ref{table:Stations}).
This facilitates calculations, and we first average the standardised CLH and HON rY indices to a common CLH-HON series which is then joined with the EKT rY indices in the same way as proceeded when reconstructing the 2st-5st long index.

The left panel of Fig. \ref{fig:Long_nonNGK_rY} shows the EKT rY index and the CLH-HON mean rY index after being standardised over the common time 1902-1925.
Overall, the two rY indices follow each other fairly closely, but there are notable differences in individual years, much larger than in the corresponding panel of Fig. \ref{fig:Long_2st_5st_rY}.
This is natural as we are comparing here indices based on one and two stations only, rather than on two and five stations where enhanced smoothing tends to reduce random scatter occurring in any individual station. 
Despite differences in individual years, the correlation between EKT and CLH-HON is good (cc = 0.87) and extremely significant (p = $2.9\cdot 10^{-8}$).

The right panel of Fig. \ref{fig:Long_nonNGK_rY} shows the non-NGK long rY index consisting of three parts, the EKT rY index before the common time (blue crosses), the average of the EKT rY index and the CLH-HON rY index during their common time (green pluses) and the CLH-HON rY index after the common time.
Again, before joining, we set all the rY levels to agree with the common-time levels and, after joining, the non-NGK long rY series was standardised over its full extent (1887-2023).

Comparing the non-NGK long rY index with the 2st-5st long rY index, one cannot see much difference.
Their mutual correlation is excellent (cc = 0.992) and extremely significant (p = $1.5\cdot 10^{-120}$).
Moreover, while the largest absolute difference between the two long indices is 0.36\,$\sigma$, the mean is only 0.1\,$\sigma$.
As already noted earlier, compared to the typical rY cycle amplitude of about 3\,$\sigma$, this difference is very small.

%
\section{Correlating long rY indices with SSN}
\label{sec:Long_SSN}

\begin{figure*}  
   \centering
  \includegraphics[width=0.98\linewidth]{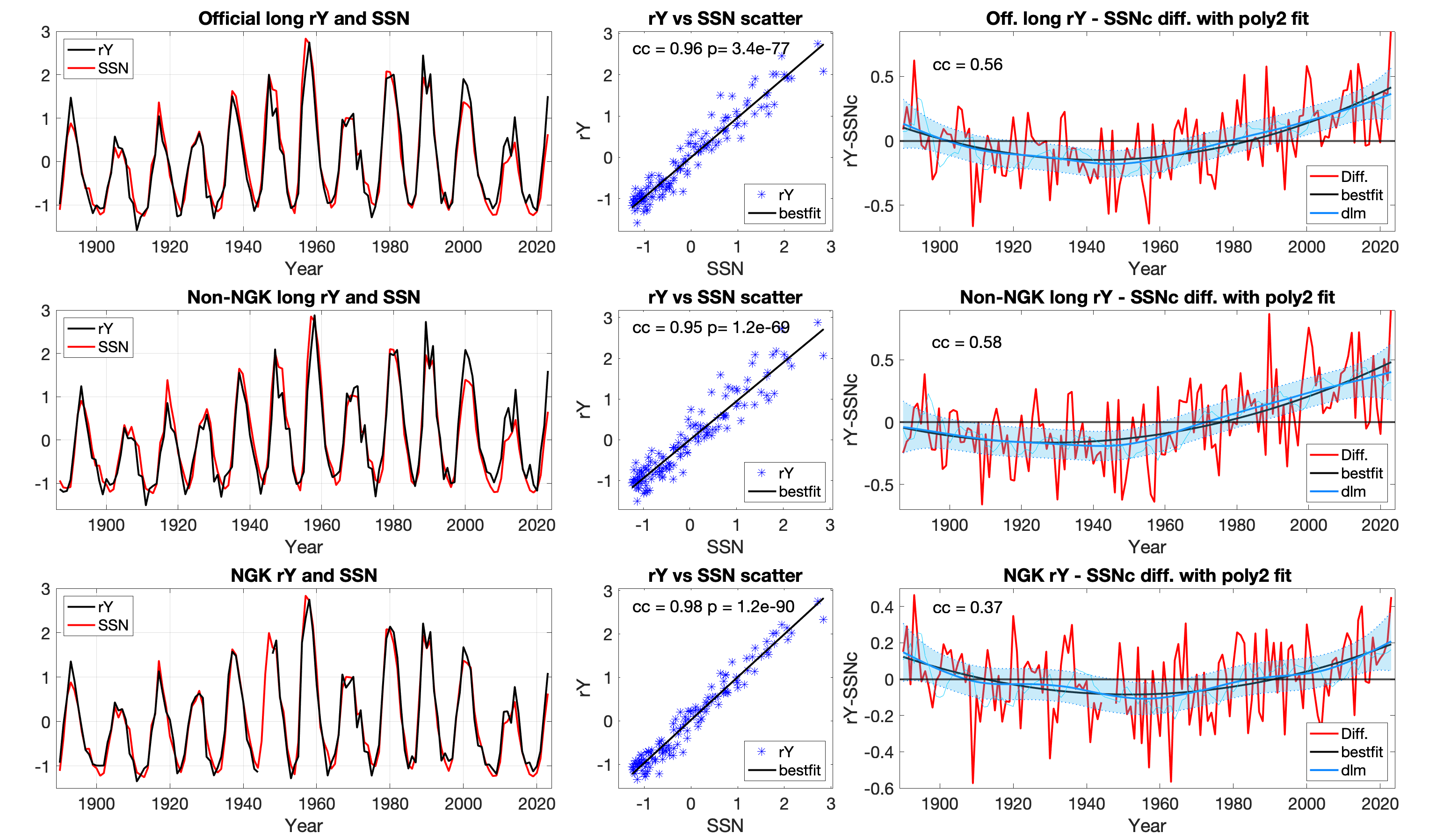}
    \caption{Three long rY series correlated with sunspot number. Presentation and panels as in Fig. \ref{fig:rY_3means_8stations_SSN_relation}.    
    Top row: official (2st-5st) rY index;
    Middle row: non-NGK rY index; 
    Bottom row: NGK rY index.
    } 
   \label{fig:rY_3Long_SSN}
\end{figure*}

The three rows of Fig. \ref{fig:rY_3Long_SSN} shows the correlation of the 2st-5st (official) long rY index (top row), the non-NGK long rY index (middle row) and the NGK rY index (bottom row) with the corresponding sunspot number series.
We have used the same form of presentation in Fig. \ref{fig:rY_3Long_SSN} as in Fig. \ref{fig:rY_3means_8stations_SSN_relation}, where the left column shows the time series of the rY index and the SSN, the middle column their scatterplot and best-fit line, and the right column the residual between the rY index and the correlated SSN.

The excellent correlation between each of the three rY indices and the corresponding sunspot number is seen in the two first columns of Fig. \ref{fig:rY_3Long_SSN}.
In each case, sunspot numbers explain more than 90\% of the variation of the yearly rY index values, and the p-values are extremely small.
This shows the close connection of the ionospheric currents and solar activity over centennial time scales.

Despite this excellent correlation, in each of the three cases, one can see the same long-term trend according to which the rY cycles are somewhat higher than the sunspot cycles for the first two cycles (SC13 and SC 14) and for the four most recent cycles (SC22-SC25) but smaller than the sunspot cycles in the intervening cycles.
This reproduces the similar observations already made for the constitutient rY series of Fig.  \ref{fig:rY_3means_8stations_SSN_relation}.

The right column of Fig. \ref{fig:rY_3Long_SSN} shows that the residual of the correlation between rY and SSN has a nonlinear temporal evolution, with relatively smaller rY values (negative residuals) occurring around 1940s - 1950s and relatively larger rY values in 1890s and 1900s and, again, during the last several decades.
A 2nd order (quadratic) polynomial fits this evolution surprisingly well (correlation coefficients are 0.5, 0.58 and 0.37 for the three cases), considering the large variability of the yearly residual values. 
The dynamic linear model 
also follows the quadratic line very closely.

A line (not shown) does not fit the data as well as the 2nd order polynomial.
The corresponding correlation coefficients are 0.34, 0.49 and 0.097 for the three cases, respectively.
Also the root mean square errors are larger for a linear fit than for a quadratic fit in each case.
However, the linear trend (increase) is statistically significant for the first two cases although less strongly than the quadratic fit.
Moreover, if we divide the time interval into two parts, both the increasing linear trend in the later part, as well as the decreasing linear fit in the earlier part would be highly significant in each case.


\begin{table}[width=.99\linewidth,cols=6,pos=h]
\caption{Correlation coefficient for 2nd order polynomial fit, and the value, error and p-value of its 2nd order term for the residuals of the three long rY series with SSN.}             
\label{table:3Long_SSN}   	
\begin{tabular*}{\tblwidth}{@{} LLLLLLL@{} }
\toprule 		
Long rY & cc & 2nd term & error & p-value \\
\midrule  		 
\textbf{Official}&0.56&$8.85\cdot 10^{-5}$&$1.45\cdot 10^{-5}$&$1.2\cdot 10^{-8}$\\  
\textbf{non-NGK}&0.58&$7.15\cdot 10^{-5}$&$1.57\cdot 10^{-5}$&$1.2\cdot 10^{-5}$\\  
\textbf{NGK}&0.37&$6.44\cdot 10^{-5}$&$1.25\cdot 10^{-5}$&$2.9\cdot 10^{-5}$\\ 
\bottomrule
\end{tabular*}
\end{table}

Table \ref{table:3Long_SSN} presents the correlation coefficients of 2nd order fits to the three residuals depicted in the right column of Fig. \ref{fig:rY_3Long_SSN}, as well as the values, errors and p-values of their second-order term coefficients.
Most importantly, the second-order term of the quadratic fit is statistically very significantly non-zero (positive) in all the three cases.
Moreover, the positive value of the second-order terms, as well as their errors, are quite similar for the three cases.
This gives strong additional evidence for the nonlinear nature of the relation between the rY index and sunspot number.

%
\section{Correlating long rY index with GSN}
\label{sec:Long_rY_GSN}

  \begin{figure*}  
   \centering
  \includegraphics[width=0.98\linewidth]{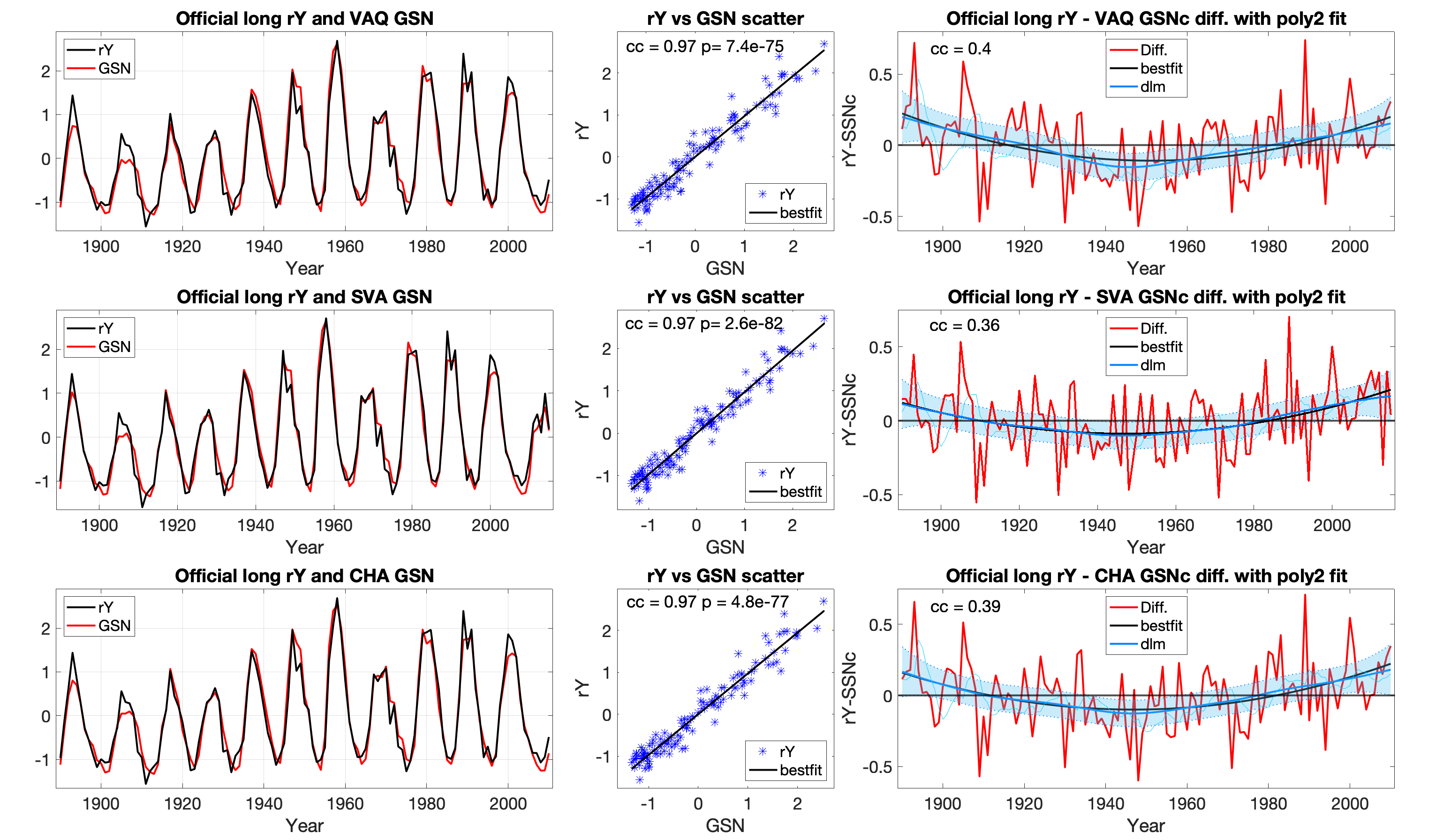}
    \caption{Official long rY series correlated with three versions of group sunspot number. Presentation and panels as in Figs. \ref{fig:rY_3means_8stations_SSN_relation} and  \ref{fig:rY_3Long_SSN}.    
    Top row: GSN series of Vaquero \cite{Vaquero_2016};
    Middle row: GSN series of Svalgaard \cite{Svalgaard_2016};
    Bottom row: GSN series of Chatzistergos \cite{Chatzistergos_2017}.
    } 
   \label{fig:rY_3Long_GSN}
\end{figure*}

The calculation method, the site and, in fact, even the name of the sunspot number from Wolf/Zürich number to International sunspot number changed in 1981 \citep{Berghmans_2006, Clette_2007}.
Therefore, one might think that the  change found in the mutual relation between sunspot numbers and the rY index index (and earlier in several other variables) could be due to a possible inhomogeneity caused by the change in the sunspot number.
Such a worry is further accentuated by recent efforts to find and correct other, earlier inhomogeneities that the sunspot number series may contain \citep{Clette_2015, Clette_Preface_2016}.

Group sunspot numbers \citep{Hoyt_SP2_1998} are another index of sunspot activity which measure the number of sunspot groups but give no weight to individual sunspots, contrary to the sunspot number.
Therefore the group sunspot number series is not affected by the changed method of sunspot number calculation in 1981. 
However, there are several GSN series which differ from each other, e.g., in the way sunspots are grouped together, but these differences have no relation to the change in SSN in 1981.
Accordingly, the group sunspot numbers can be used to test the validity of the observed change in the mutual relation of photospheric (sunspots) and chromospheric (rY, EUV) activity. 

As representative GSN series we use here the collection of sunspot groups from 1610 to 2010 by \cite{Vaquero_2016} (to be called here VAQ GSN), which is a revision of the original collection of sunspot groups by \cite{Hoyt_SP2_1998}, the collection of sunspot groups from 1610 to 2015 by \cite{Svalgaard_2016} (SVA GSN) and the collection of sunspot groups from 1739 to 2010 by \cite{Chatzistergos_2017} (CHA GSN).

Figure \ref{fig:rY_3Long_GSN} depicts the relation between the official long rY index and these three GSN series in the same format as Figs. \ref{fig:rY_3means_8stations_SSN_relation} and  \ref{fig:rY_3Long_SSN}.
The left column of Fig. \ref{fig:rY_3Long_GSN} shows that the rY cycle peaks are higher than GSN cycle peaks for the first two cycles (SC13-SC14) and for the last two-three cycles (SC22-SC24; Note that all GSN series are shorter than SSN, all missing SC25 and two of them even SC24).
This depicts a similar trend as the corresponding relation between rY and SSN except that for GSN the difference between peaks is larger, especially for SC14. 
For the intervening cycles, GSN cycles are not systematically higher than the rY cycles, but GSN cycle minima tend to attain higher levels than rY cycle minima, especially for cycles 14-17, similarly as for SSN.
The middle column of Fig. \ref{fig:rY_3Long_GSN} shows that the GSN correlates with rY equally well for all GSN versions and even slightly better than the same index correlates with SSN (see top middle panel of Fig. \ref{fig:rY_3Long_SSN}).


\begin{table}[width=.99\linewidth,cols=6,pos=h]
\caption{Correlation coefficient for 2nd order polynomial fit, and the value, error and p-value of its 2nd order term for the residuals of the long rY series with three GSN series.}             
\label{table:corr_3Long_GSN}   	
\begin{tabular*}{\tblwidth}{@{} LLLLLLL@{} }
\toprule 		
Long rY & cc & 2nd term & error & p-value \\
\midrule  		 
\textbf{VAQ GSN}&0.40&$8.88\cdot 10^{-5}$&$1.87\cdot 10^{-5}$&$0.62\cdot 10^{-5}$\\  
\textbf{SVA GSN}&0.36&$6.42\cdot 10^{-5}$&$1.59\cdot 10^{-5}$&$9.47\cdot 10^{-5}$\\  
\textbf{CHA GSN}&0.39&$8.05\cdot 10^{-5}$&$1.81\cdot 10^{-5}$&$1.97\cdot 10^{-5}$\\ 
\bottomrule  	
\end{tabular*}
\end{table}

The right column of Fig. \ref{fig:rY_3Long_GSN} depicts the residuals of the correlation between the long rY index and the three GSN series in the same format as in Figs. \ref{fig:rY_3means_8stations_SSN_relation} and \ref{fig:rY_3Long_SSN}.    
All the three GSN series show a nonlinear evolution with the long rY index, quite similarly as SSN with the three long rY indices (see Fig. \ref{fig:rY_3Long_SSN}).
 We have listed in Table \ref{table:corr_3Long_GSN} the correlation coefficients of the 2nd order polynomial fits depicted in Fig. \ref{fig:rY_3Long_GSN}, and the related values, errors and p-values of the 2nd order term for each case.
Note that not only the correlation coefficients for the three cases are very similar (varying from 0.36 to 0.40) but also all the parameters related to the 2nd order term are quite similar for the three cases.

Most importantly, Table \ref{table:corr_3Long_GSN} shows that in each case the 2nd order term of the polynomial is very significantly non-zero (positive) with all p-values being smaller than 0.0001 (99.99\% confidence level). 
Note also that all the values of the 2nd-order term for the three GSN relations given in Table \ref{table:corr_3Long_GSN} and for the three SSN relations listed in Table \ref{table:3Long_SSN} are quite similar.
Even the corresponding errors and p-values are of the same order of magnitude.
This gives strong evidence that the nonlinear relation found between the sunspot number and the rY index is not due to a possible inhomogeneity of sunspot number  series but a general, physical property of the long-term relation between solar parameters describing photospheric activity and parameters describing activity at higher altitudes (chromosphere and low corona) of the solar atmosphere.

%
\section{Correlating long rY index with EUV proxies}
\label{sec:Long_rY_EUV}

\begin{figure*}  
   \centering
  \includegraphics[width=0.98\linewidth]{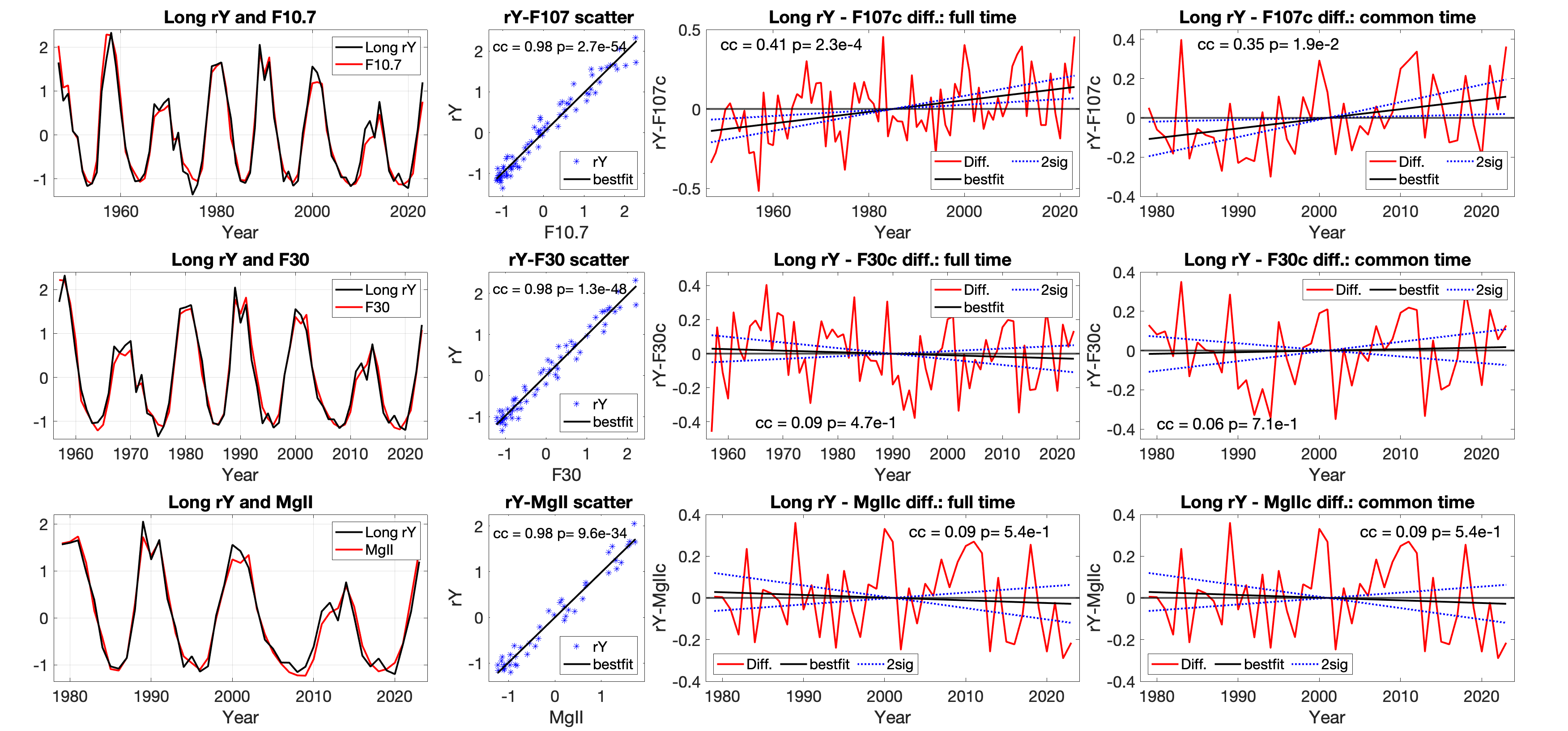}
    \caption{
    Correlating the official long rY index with three EUV proxies: F10.7 (top row), F30 (middle row) and MgII index (bottom row).
    Left column: Time series of long rY and EUV proxy;
    Second column: Their scatterplot with best-fit line;
    Third column: Residual of correlation between the long rY index with three EUV proxies over the whole time interval, together with the best-fit line (black) and two lines with slopes two standard deviations above and below the best-fit line slope (blue dotted lines);
Right column: Same as in the third column but over a shorter time interval common to all EUV proxies (1979-2023).
    } 
   \label{fig:Long_rY_EUV}
\end{figure*}

In this Section we will compare the official long rY index with three other EUV proxies, the two radio fluxes at 10.7\,cm and 30\,cm, and the Bremen MgII index.
We have correlated the long rY index with each of these three EUV proxies over the whole temporal extent of each proxy (1947-2023 for F10.7, 1957-2023 for F30 and 1979-2023 for MgII) and presented the time series, scatterplots and residuals as the first three columns in Fig. \ref{fig:Long_rY_EUV}. 

The residual plots include, in addition to the best-fit line (black), two lines with slopes that are two standard deviations above or below the best-fit line slope (blue dotted lines). 
In addition, in order to better compare the evolution of the three proxies, we have correlated the long rY with the same three proxies over a shorter  time interval common to all three proxies (1979-2023; dictated by the shortest index, MgII), and presented the corresponding residuals as the right-most column in Fig. \ref{fig:Long_rY_EUV}.
(Note that the two residuals in the bottom row of Fig. \ref{fig:Long_rY_EUV} are the same).

The two left-most columns of Fig. \ref{fig:Long_rY_EUV} show that all the three EUV proxies have the same correlation coefficient (0.98) with the long rY index. 
This is very convincing for the validation of the rY index as a long-term proxy of solar EUV irradiance.
Correlation remains equally good for the shorter (last 45 years) and for the longer (last 77 years) time interval.
In all cases, statistical significance is extremely good.
Note that the p-values are clearly organised by the length of the EUV proxy, with the longest proxy leading to much smaller p-value (F10.7 with $2.7\cdot 10^{-54}$ than the shortest proxy (MgII with $9.6\cdot 10^{-34}$).
This proves that the long rY index is an excellent, homogeneous EUV proxy over different time intervals.

The two right-most columns of Fig. \ref{fig:Long_rY_EUV} show that, despite the same level of correlation between the long rY index and the three different EUV proxies, the three EUV proxies have somewhat different long-term relations with the rY index.
Starting from the bottom row of Fig. \ref{fig:Long_rY_EUV}, the MgII index shows no significant trend with respect to the rY index over its full range (1979-2023). 
The best-fit line does not explain much of the variability of the residual (cc = 0.09), and its slope does not significantly deviate from zero (p = 0.54).
The x-axis (line with slope = 0) almost overlaps with the best-fit line and remains well between the two blue lines with slopes two standard deviations above and below the best-fit line slope.
Since the MgII index is the standard measure of solar EUV activity, we can conclude that the long rY index not only correlates with the MgII index extremely well but also reproduces its long-term trend reliably over the full time range of the MgII index (1979-2023).
This verifies that the rY index is a very reliable index of solar EUV variability.

The residual of the correlation of the rY index with the F30 radio flux (middle row of Fig. \ref{fig:Long_rY_EUV}) repeats the same observation as for the MgII index.
The F30 flux shows no trend with respect to the rY index, not over its full range (1957-2023; pp = 0.47), nor over the shorter time interval (1979-2023; p = 0.71).
Accordingly, the rY index can be used as a reliable proxy of solar EUV variability even before the MgII time interval, at least from the late 1950s onward.

The top row of Fig. \ref{fig:Long_rY_EUV} shows that the rY index has a slightly increasing trend with respect to the F10.7 radio flux.
Over the shorter time interval (1979-2023; top right panel of Fig. \ref{fig:Long_rY_EUV}) the trend is only marginally nonzero (p = 0.019) and explains only 12\% of the residual variability.
However, over the full extent of the F10.7 flux the increasing trend is more significant (p = $2.3\cdot 10^{-4}$) and explains 17\% of variability.
We note that the non-parametric Mann-Kendall test \citep{Mann_1945, Kendall_1975} yields quite similar estimates for the p-value (p = 0.026 for short and p = $7.9\cdot 10^{-4}$ for full series) as the normal least-squares method. 
Moreover, the least-squares fit and the non-parametric method by \citet{Sen_1968} (median of all pairwise slopes) give very similar slope estimates (0.0050 and 0.0049 for short and 0.0036 and 0.0034 for full series, respectively).

The increasing trend of the rY index with respect to the F10.7 flux agrees with the earlier finding \citep{Mursula_AA_2024} that the F30 radio flux, as well as F15, a third radio flux at 15\,cm wavelength, have a statistically significant increasing trend with respect to F10.7.
We have also verified that, as expected from Fig. \ref{fig:Long_rY_EUV}, both F30 flux and MgII index have a significant increasing trend with respect to F10.7 flux similarly as the rY index.
We note that other studies \citep{Lastovicka_2021a, Lastovicka_SpW_2023} have found that the ionospheric parameters like the foF2 and foE indices (maximum densities in the upper and lower ionosphere, respectively) fit better with the F30 flux than F10.7 flux.
Accordingly, the F10.7 radio flux may not be the ideal proxy for solar EUV irradiance as it lacks some of the long-term increase that the other, more reliable EUV proxies (MgII, F30, rY) include.

%
\section{Discussion}
\label{sec:Discussion}

\subsection{Method}
\label{sec:Disc:Method}

We have used here eight early and long-operating observatories with reliable, homogeneous geomagnetic measurements covering 137 years from 1887 to 2023.
We calculated the yearly means (rY indices) of the amplitude (range) of the daily variation of the magnetic Y (East-West)-component, which were first used as a proxy of sunspot activity \citep{Lamont_1851,Gautier_1852, Wolf_1852} and then, more recently and more correctly, as a proxy of solar EUV irradiance \citep[see, e.g., ][]{Svalgaard_EUV_2016}.

It is known that solar EUV irradiance produces and controls the intensity of a dayside ionospheric current system called the Sq current, which consists of two oppositely flowing current vortices around the sub-solar point.
The intensity of the Sq current system and, thereby, of solar EUV irradiance can be robustly measured from the amplitude of the local time variation of the Y-component, which is closely similar at all low- and mid-latitude stations in one hemisphere.

The rY method has a few benefits that make it reliable and homogeneous over long time scales.
The daily amplitude is rather insensitive to solar flares, which produce random deflections to the daily curve that average out in the yearly means.
Moreover, the yearly rY indices are practically the same for all days and only quiet days \citep{Chree_1913}, which implies that they are not affected by geomagnetic activity (particle precipitation into the upper atmosphere).
Accordingly, the rY method measures the long-term variation of the EUV baseline, and is not affected by short-term eruptions producing temporary enhancements in solar electromagnetic radiation or particle precipitation.

The rY method also has some important benefits compared to other methods of measuring EUV irradiance.
Geomagnetic measurements remain homogeneous over long time scales since they experience little or no temporal degradation contrary, e.g., to satellite EUV instruments that suffer from radiation damage caused by the measured EUV and other hard photons and solar energetic particles.
On the other hand, ground-based solar spectral line observations, such as chromospheric CaII K-line measurements, do cover a century of EUV measurements but suffer from insufficient or missing documentation, changes in instrumentation and detection method, uneven data sampling and from varying atmospheric transparency.
Accordingly, we are confident that the current rY method gives the longest homogeneous estimate of solar EUV irradiance, most likely the only reliable estimate that covers the whole Modern Maximum.

One might think that our method of using linear correlations between the different variables could affect or even cause the observed drifts.
Although the daily values of practically all photospheric and chromospheric parameters have at least a slightly nonlinear relation \citep[see, e.g.,][]{Tapping_2017, Clette_F107_2021,Yeo_2020}, the corresponding yearly means are almost purely linearly related \citep{Foukal_1998, Mursula_AA_2024}.
The suggested cause of such short-term nonlinearity is the different lifetime of (short, a few days) sunspots and (long, several weeks) plages \citep[see, e.g.,][]{Preminger_2007} which, however, is alleviated at time scales longer than either lifetime.
We have earlier compared the linear and (weakly) nonlinear fits between sunspots and radio fluxes and shown that the two fits yield almost exactly the same temporally changing relation during the last 60 years \citep{Mursula_AA_2024}.
(The same roughly linear increase in the relation between rY and sunspots is seen for the corresponding part of Fig. \ref{fig:rY_3Long_SSN}).
Accordingly, the temporal change in the relation between the different solar variables is not an artifact of using linear relations but a genuine physical change between photospheric and chromospheric variables, which even has tangible effects to the Earth (see later).

\subsection{Solar EUV proxies}
\label{sec:Disc:EUV}

We showed that, although the ranges of secular variation of the magnetic Y-component can be very large and very different at the different observatory sites, the yearly rY indices attain quite similar absolute levels and follow the cyclic variation and even the long-term trend very similarly for all stations.
This gives the possibility to combine rY values observed at different sites in order to have the longest possible proxy for solar EUV irradiance.
Using the extended station network we have derived two long multi-station rY indices which, together with the NGK rY series cover the time from 1890s onward, including the last decades of the 19th century, the weak cycles around the turn of the 19th and 20th centuries, all of the Modern Maximum, and even the ongoing period of weak solar activity after the Modern Maximum.

Three long rY indices were constructed from different selections of stations in order to have as versatile and independent estimates as possible.
They all agree on the fact that the relation between the rY index and sunspot number is nonlinear over the studied time interval (1887-2023).
The relative fraction of sunspots is larger than the rY index (EUV irradiance) during the growth and maximum phase of the Modern Maximum, when the overall solar activity and cycle heights are increasing and large, while the reverse is true when the overall activity is decreasing or weak during the decay of the Modern Maximum and during the decay of the 19th century maximum.

This nonlinear relation is not due to possible inhomogeneities in the sunspot number.
We used three different group sunspot number series \citep{Vaquero_2016, Svalgaard_2016, Chatzistergos_2017} to show that quite a similar, significantly nonlinear relation exists between the long rY index and all of the three GSN series.
This suggests that the nonlinearity is generally valid between several types of photospheric variables (sunspots) and variables related to solar activity at higher altitudes of the solar atmosphere (chromosphere and low corona).

We have shown that the rY index correlates extremely well with the standard EUV proxy, the MgII core-to-wing ratio index.
Moreover, these two parameters do not drift with respect to each other but, rather, have the same trend over the whole common time interval from the late 1970s onward.
This verifies that the rY index is a reliable, homogeneous long-term index of solar EUV irradiance.
This is also in agreement with the earlier finding that the MgII index increases with respect to the sunspot number \citep[see Fig. 6 of ][]{Mursula_AA_2024}. 

Earlier studies show that the F10.7 radio flux, which is often used as an extended proxy of EUV irradiance, also increases with respect to sunspots during the last 70 years \citep{Mursula_AA_2024}.
However, the rY index increases slightly but significantly also with respect to the F10.7 radio flux over this  time.
Also the MgII index increases with respect to the F10.7 radio flux over their common time.
This implies that, although the F10.7 flux correlates very well with MgII and rY, it does not have exactly the same long-term trend and, therefore, cannot be considered a reliable long-term index of EUV irradiance.
Note that this is not due to a possible inhomogeneity in the F10.7 index series, as earlier suggested by \cite{Clette_F107_2021}, but reflects a genuine difference in how the different layers of solar atmosphere evolve with varying activity \citep{Mursula_AA_2024}.

The F30 radio flux correlates extremely well with rY and MgII, and also has the same long-term trend over the common time intervals (from 1950s onward with rY).
Therefore, the F30 flux is a reliable long-term proxy of EUV irradiance, rather than the F10.7 flux.
This agrees with 
the earlier finding that the F30 increases with respect to F10.7 \citep{Mursula_AA_2024}.
Note also that this trend difference gives evidence for a long-term change in solar spectrum at radio frequencies.


Our results are in an excellent agreement with the observation \citep{Lastovicka_SpW_2023} that ionospheric critical frequencies (foF2 and foE; electron density) observed at several stations increased from 1976-2014 relatively larger than predicted either by F10.7 or sunspots, but were in an agreement with F30.
Moreover, it is well established \citep{Maruyama_2010, Goncharenko_2021} that MgII outperforms F10.7 and sunspot number in explaining the long-term change in the total electron content (TEC).
Based on these results it has been suggested \citep{Lastovicka_SpW_2023, Qian_Mursula_2025} that F30 should be used to describe the ionospheric response rather than the conventionally used F10.7.
Our results allow us to suggest that the long rY index can be used for the same purpose over much longer times from 1890 onward.
These considerations show that there are practical consequences from the observed long-term change in the relation between the different solar variables.

\subsection{Implications for solar atmosphere}
\label{sec:Disc:Implications}

Solar EUV irradiance mainly originates in the chromosphere and lower corona above active regions called plages \citep{Pallavicini_1981, Neupert_1998}.
Plages are wide active regions of moderately intense magnetic field which may or may not include sunspots that have much stronger magnetic fields than an average plage field.
Sunspots typically form only a small fraction of the total plage area.
Solar radio fluxes mainly originate from plages, where they are typically produced by electron bremsstrahlung, but a part of radio flux is emitted in the strong magnetic fields of sunspots via cyclotron resonance \citep{Schonfeld_2019}.

The changing long-term relation between sunspots and EUV irradiance suggests that the relation between photospheric sunspots and chromospheric plages changes when the overall solar activity varies over the Gleissberg cycle.
During the growth and maxima of Gleissberg cycles, sunspots increase relative to plages and, during the decay, plages increase relative to sunspots.
The fact that the F10.7 flux increases less rapidly with respect to sunspots than MgII and rY indices may be related to the fact that a fraction of the F10.7 flux is produced in sunspots.


We have suggested earlier \citep{Mursula_AA_2024} that the physical reason to the observed long-term changes is that the magnetic structure and, as a consequence, the plasma profile of the solar atmosphere change with long-term solar activity.
As solar activity increases (or weakens) during the growth (decay) of a Gleissberg cycle (like the Modern Maximum), the parameters describing magnetic field elements of different field strength (in particular, sunspots and plages) or being produced at different heights of solar atmosphere (e.g., F10.7 and F30 fluxes), 
vary differently.
With increasing activity, there will be more of strong magnetic flux tubes, which makes the typical magnetic field somewhat more radial, thereby raising the magnetic canopy structure higher and extending the magnetic field lines horizontally further out from their footpoint \citep{Virtanen_ApJL_2020}.
This will reduces the relative fraction of plages compared to sunspots.

In the opposite phase, decreasing long-term activity will lower the magnetic canopy structure and modify the plasma density profile. 
Due to decreasing plasma density, the sources of solar radio waves will move to lower altitudes to adjust to the plasma density corresponding to their respective wavelength (bremsstrahlung frequency depends on electron density).
However, the longer radio waves have to move a relatively smaller step down than shorter radio waves because of a larger volume compression (density increase) at a higher altitude in the canopy structure of magnetic field lines.
Moving downward to a cooler temperature in solar atmosphere all radio emissions reduce, but the longer waves cool relatively less than the shorter waves, which explains the increase of longer waves relative to shorter waves when activity is decreasing.

\subsection{Quantifying the change over MM}
\label{sec:Disc:Quantifying}

We have earlier estimated \citep[see][Fig. 6 and Sec. 7]{Mursula_AA_2024} that the increase of the MgII index with respect to sunspots in 1979-2021 is about 15\% of the solar cycle variability (cycle maximum - minimum difference) of MgII.
Using the longer rY index and its excellent correlation with the MgII index, we can now estimate the total increase of MgII vs SSN over the whole Modern Maximum, including the nonlinear part of this relation in addition to the later, quasi-linear part.
(Note here the virtue of standardisation which allows direct comparison between different standardised variables).

Using the best-fit polynomial of Fig. \ref{fig:rY_3Long_SSN} we find that the increase of rY during the roughly linear part from 1979 to 2021 is about 0.41$\sigma$.
This part corresponds to the 15\% fraction of MgII cycle variability.
The nonlinear part from the minimum in 1940s to 1979 gives an additional increase of about 0.25$\sigma$.
Accordingly, the total increase of the MgII index (EUV irradiance) with respect to sunspots during the decay of the Modern Maximum was about 0.41$\sigma$ + 0.25$\sigma$ = 0.66$\sigma$.
This corresponds to a fraction of 0.66/0.41*15\% = 24\% from typical MgII cycle variability.
This is a significant change and, should is be applicable to other forms of EUV irradiances or, perhaps, even to a large part of solar spectrum, it could imply considerable changes to the estimated long-term change of many solar variables, including solar total irradiance.

\subsection{Implications for solar and stellar evolution}
\label{sec:Disc:Evolution}

The magnetic fields of (chromospheric) plages are rooted in the photosphere where they are seen as bright faculae. 
Faculae, as plages, are regions of enhanced magnetic fields but much weaker than sunspots.
Although the long-term variation of faculae is less well known, it is commonly assumed that faculae follow the evolution of plages \citep[see, e.g.,][]{Chatzistergos_SWSC_2020, Yeo_2020}.
Because faculae are brighter than the average surface, they are important structures for solar brightness (total solar irradiance, TSI) which is formed by the competing balance between darkening caused by sunspots and brightening produced by faculae \citep{Hall_2009, LockwoodGW_2007, Radick_2018}.

Our results indicate that the sunspot-facula ratio varies between the growth/ maximum and the decay phases of the Modern Maximum and, very likely, between the corresponding phases of other Gleissberg cycles and other similar large variations of solar activity. 
It is also quite likely that the same consideration applies to large activity variations of Sun-like stars.
The varying sun/starspot-facula ratio will also modify the estimated evolution of solar and stellar brightness. 

Young Sun-like stars rotate fast and create very strong magnetic fields with large sunspots which dominate brightness variation and make brightness decrease with stellar magnetic activity (out-of-phase relation between brightness and magnetic activity).
In a later phase of stellar evolution the magnetic fields are much weaker but still strong enough to create some number of spots, although the faculae dominate brightness variation and then produce an in-phase variation of brightness with activity.
The Sun is now in this later phase of its stellar evolution, as verified by the observed  in-phase relation of its brightness and magnetic activity \citep{Willson_1988, Frohlich_2004, Frohlich_2013, Kopp_2016}.

The effect of the varying starspot-facula ratio will be largest for those stars that are close to the transition from the starspot-dominated period to facula-dominated period.
In fact, the Sun itself has relatively recently passed this transition in its stellar evolution \citep{Hall_2009, LockwoodGW_2007, Reinhold_2019}. 
Therefore, the balance between the two competing factors is particularly delicate for the current Sun.
Indeed, this is also the likely reason why this variability is recognisable in the current Sun, even at a considerable level.

Our results suggest that the starspot-facula ratio varies with stellar activity for Sun-like stars.
For stars close to the transition from a spot-dominated phase to a facula-dominated phase, this could modify the currently estimated relation between brightness and activity. 
It may even turn out that sufficiently long observations of brightness and activity variations of a Sun-like star could reveal how close the star is to this transition.
This and other stellar implications will have to wait for a dedicated study.

%
\section{Conclusions}
\label{Sec: Conclusions}

Using different combinations of magnetic data from eight long-operating observatories, and the method of daily variation of declination (Y-component) introduced by the early explorers \citep{Lamont_1851, Gautier_1852, Wolf_1852}, we have extracted three versions of the rY index, a yearly-resolved proxy of solar EUV irradiance covering 137 years from 1887 to 2023.
All the three long rY series give the same main results on the long-term relations between the rY index and several solar variables included in this study.

We have shown that the rY index correlates extremely well with the MgII index and the solar F30 radio flux.
These three indices have no trend relative to each other over their whole respective intervals (1979-2023 and 1957-2023).
On the other hand, we find that the solar F10.7 radio flux has a significant long-term trend with respect to the three co-varying EUV indices (MgII, F30, rY).
Therefore, the rY index replaces the F10.7 index as the long-term EUV proxy, and extends the MgII index by 90 years.

We have verified here the earlier observation that, during the last 50-70 years, all the four studied EUV proxies (rY, MgII, F30, F10.7) have an increasing trend with respect to sunspots, irrespective of whether they are counted as Wolf numbers or group numbers \citep{Mursula_AA_2024, Mursula_AA_2025}.
Moreover, using three versions of the centennial rY index, we have verified that the long-term relation between EUV irradiance and sunspots has a nonlinear, close to quadratic, temporal evolution over the Modern Maximum \citep{Mursula_AA_2025}.

We find that, due to this centennial nonlinear change in the relation between the rY index (chromosphere) and sunspots (photosphere), the Sun has more sunspots relative to EUV irradiance during the growth and maximum of the Modern Maximum, while the opposite is true during its recent decay.
We estimate that the MgII index (solar EUV irradiance) increases by 24\% of its solar cycle variation with respect to the sunspot number from the minimum during the last 70 years.

These results indicate a systematic difference in the long-term evolution between sunspots (generally: photosphere) and plages (generally: chromosphere), which is changing with long-term solar activity.
This change may be visible because of the close proximity of the Sun to the transition from a spot-dominated to facula-dominated phase in its stellar evolution.
It is likely that similar changes take place in Sun-like stars sufficiently close to the transition.
Thus, our results may affect the evolution of stellar brightness and the interpretation of the long-term observations of Sun-like stars.


\section{Acknowledgements}
%

The magnetic data were provided by the World Data Center for Geomagnetism at the British Geological Survey (https:
//wdc-dataportal.bgs.ac.uk). 
The sunspot numbers (version 2; https://www.sidc.be/SILSO/datafiles) and the three collections of the number of sunspot groups (https://www.sidc.be
/SILSO/groupnumberv3) were provided by the World Data Center SILSO.
The NOAA F10.7\,cm index was obtained from the LISIRD server (https://lasp.colorado.edu/lisird/).
The recent Penticton 10.7\,cm data were retrieved from the NRCan server (https://www.spaceweather.gc.ca/forecast-prevision/
solar-solaire/solarflux/sx-5-flux-en.php), which is served by the Solar Radio Monitoring Program operated jointly by the National Research Council Canada and Natural Resources Canada with support from the Canadian Space Agency. 
The dynamic linear model of Marko Laine can be found at https:
//mjlaine.github.io/dlm/index.html.


\bibliography{rY_SSN_FINAL}

\end{document}